\documentclass[twocolumn,10pt,amsmath,amssymb,superscriptaddress,pra,aps]{revtex4-1}
\usepackage{epsfig}
\usepackage{color}
\usepackage{graphicx}
\usepackage{bm}
\usepackage{epstopdf}
\usepackage[T1]{fontenc}

\begin{document}
\title{Four fermions in a one-dimensional harmonic trap: Accuracy of a variational-ansatz approach}

\author{D. \surname{P{\k e}cak}}
\affiliation{\mbox{Institute of Physics, Polish Academy of Sciences, Aleja Lotnikow 32/46, PL-02668 Warsaw, Poland}}
\author{A.~S. \surname{Dehkharghani}}
\affiliation{Department of Physics and Astronomy, Aarhus University, DK-8000 Aarhus C, Denmark}
\author{N.~T. \surname{Zinner}}
\affiliation{Department of Physics and Astronomy, Aarhus University, DK-8000 Aarhus C, Denmark}
\author{T. \surname{Sowi\'{n}ski}}
\affiliation{\mbox{Institute of Physics, Polish Academy of Sciences, Aleja Lotnikow 32/46, PL-02668 Warsaw, Poland}}

\date{\today}

\begin{abstract}
Detailed analysis of the system of four interacting ultra-cold fermions confined in a one-dimensional harmonic trap is performed. The analysis is done in the framework of a simple variational ansatz for the many-body ground state and its predictions are confronted with the results of numerically exact diagonalization of the many-body Hamiltonian. Short discussion on the role of the quantum statistics, i.e. Bose-Bose and Bose-Fermi mixtures is also presented. It is concluded that the variational ansatz, although seemed to be oversimplified, gives surprisingly good predictions of many different quantities for mixtures of equal as well as different mass systems. The result may have some experimental importance since it gives quite simple and validated method for describing experimental outputs.
\end{abstract}

\maketitle

\section{Introduction}
One-dimensional systems of few quantum particles have attracted a lot of attention in the past few years due to the amazing experimental progress in studying such systems. At last, it becomes possible not only to test and improve theoretical description of such systems \cite{lieb1966mathematical,baxter1989exactly,Giamarchi2003,plischke2006equilibrium,GriningNJoP2015,DecampAe2016,VolosnievNC2014,Dobrzyniecki2016DoubleWell,Koscik2012,Koscik2015,Koscik2017,DeuretzbacherPRA2014,YangPRA2015,YangPRA2016,GharashiPRL2013,Garcia-MarchNJoP2014}, but also to test all these theoretical ideas experimentally \cite{blochRMP2008,lewensteinAiP2007,esslingerARoCMP2010,moritzPRL2003,stoferlePRL2003,KinoshitaS2004,KinoshitaN2006,ParedesN2004,HallerS2009,haller2010pinning,PaganoNP2014}. New experiments of an extremely high accuracy have challenged theoreticians to serve predictions with incredible precision and as a consequence to audit previous rough approximations made to describe  properties of few quantum bodies \cite{MurmannPRL2015a,MurmannPRL2015b,SerwaneS2011,WenzS2013,zurnPRL2013,zurnPRL2012,Kaufman2015Entangling}.

The physics of few quantum particles is extremely difficult to be analyzed without any approximations. It comes from the simple observation that 'a few' is too many to use a straightforward method for one- and two-body physics, and at the same time it is still not enough to adopt methods of statistical many-body theory and mean-field description \cite{GriningPRA2015,Schmitz2013Breathing,Blume2010TwoThreeToMany}. Therefore, one has to find completely different approaches to the problem (for example those, which were up to now in a domain of nuclear physics \cite{zinner-jensen2013}). Independently of these facts, there always exists a temptation to describe complicated few-body problem with evidently oversimplified methods. One of these kind of approaches is based on different implementations of the variational-ansatz method.

In this paper we want to investigate the properties of a system of four fermionic atoms confined in a one-dimensional harmonic trap obtained via a simple variational method and validate these results. The method is based on an assumption that the ground state of a many-body interacting system can be almost perfectly superposed from two limiting many-body states, i.e., the ground states obtained for vanishing and very strong repulsions \cite{AndersenAe2016}.
Since the method was successfully adopted for systems of two and three quantum particles (and for a particular class of polaron systems with up to six bodies), a natural question about the validity of this assumption for larger number of particles arises with other system compositions. Here we try to answer this question by comparing predictions of the ansatz with predictions of numerically exact diagonalization of the four-body Hamiltonian. A comparison is done on various levels by considering many different quantities that, in principle, may be extracted from the experimental data. We stress that we consider the experimentally relevant situation where the particles in our system have different masses. This is a particularly difficult issue for one-dimensional systems. Such systems cannot be addressed using for instance the Bethe ansatz as mass differences will generically break the assumption of non-diffractive scattering. This assumption is central to the traditional Bethe ansatz approach to generate exact solutions of one-dimensional many-body systems \cite{baxter1989exactly,Giamarchi2003}. This implies that a simple approach to mass imbalanced systems is highly desirable.

In Sec.~II a brief description of the system under study is given and both complementary methods of treatment, i.e., the interpolatory ansatz and the exact diagonalization are briefly characterized. In Sec.~III we compare different predictions of both methods and we discuss disclosed discrepancies. Finally, in Sec.~IV we give some remarks on four-body systems with other quantum statistics, we discuss some possible extensions of the variational method and conclude briefly.

\section{The model}
{\bf The system studied.}---
We consider $N_a~=~N_b~=~2$ fermionic particles confined in an external one-dimensional harmonic potential of frequency $\omega$.
In principle, the particles of different kinds may have different masses, i.e., $m_a\neq m_b$. We assume that interactions between particles can be described with two-body contact $\delta$-like potential. In this case, due to the fermionic nature of particles, the interactions are present only between particles of different components.
The Hamiltonian of the system has the form
\begin{align} \label{Ham}
\mathcal{H}&=\mathcal{H}_a+\mathcal{H}_b+\mathcal{H}_{ab}, \\
\mathcal{H}_a&=-\frac{\hbar^2}{2m_a}\left(\frac{\partial^2}{\partial q_1^2}+\frac{\partial^2}{\partial q_2^2}\right)+\frac{m_a\omega^2}{2}\left(q_1^2+q_2^2\right), \nonumber \\
\mathcal{H}_b&=-\frac{\hbar^2}{2m_b}\left(\frac{\partial^2}{\partial q_3^2}+\frac{\partial^2}{\partial q_4^2}\right)+\frac{m_b\omega^2}{2}\left(q_3^2+q_4^2\right), \nonumber \\
\mathcal{H}_{ab}&=g\left[\delta(q_1\!-\!q_3)+\delta(q_2\!-\!q_3)+\delta(q_1\!-\!q_4)+\delta(q_2\!-\!q_4)\right]. \nonumber
\end{align}
Notice that since the masses are allowed to be different, the Bose-Fermi mapping \cite{girardeauJoMP1960,girardeauPRA2004,girardeauPRL2007} cannot be applied. However, in extreme limits the exact ground-state wave function is always known. For vanishing interactions, $g=0$, particles occupy the two lowest single-particle orbitals of corresponding harmonic oscillators. In the case of infinitely strong interactions, $1/g=0$,
a semi-analytical expression for the exact four-body ground state was found recently in \cite{DehkharghaniJoPBAMaOP2016} using the methods introduced in \cite{DehkharghaniSR2015,DehkharghaniPRA2015}. In general, the properties of the ground state for intermediate interactions cannot be found analytically and one needs to use numerical or approximate methods.

{\bf Interpolatory ansatz.}---
Quite recently, it was proposed to use a very simple variational method based on the assumption that the ground state of the system for any interaction can be well approximated by an appropriate superposition of the ground states in the limiting cases:
\begin{equation} \label{variat}
|\Psi(g)\rangle = \alpha(g)|\Psi_0\rangle + \beta(g)|\Psi_\infty\rangle.
\end{equation}
The coefficients $\alpha(g)$ and $\beta(g)$ are determined by minimizing an expectation value of the many-body Hamiltonian \eqref{Ham} in this state. Note, that the many-body states $|\Psi_0\rangle$ and $|\Psi_\infty\rangle$ are not necessarily orthogonal. Therefore, the variational parameters fulfill non-natural normalization conditions. The detailed prescription for obtaining appropriate variational parameters $\alpha(g)$ and $\beta(g)$ was discussed in \cite{AndersenAe2016}. For the completeness of our discussion we include a brief discussion of the method in the Appendix. A small modification of the method also mentioned in the Appendix which substantially improves predictions of the ground-state energy is discussed in further analysis.

Although the ansatz seems to be highly oversimplified it was used for systems with equal masses with surprisingly good results.
Here we want to make a comprehensive study of the  accuracy of the ansatz when different quantities and interparticle correlations extracted from the ground state are considered. Especially, we are interested in the cases when the particles belonging to the different components have different masses. To find quantitative answers to this open questions we perform the numerically exact diagonalization of the many-body Hamiltonian Eq.~(\ref{Ham}), we find its exact ground state as a function of interactions and we compare different quantities with predictions of the variational ansatz.

{\bf Numerical diagonalization.}---
The exact diagonalization is performed in a straightforward and well-established way. First, we express the many-body Hamiltonian Eq.~\eqref{Ham} in a matrix form in an appropriate Fock basis. It can be done by expressing all many-body states of the system in the basis composed as products of single-particle orbitals:
\begin{equation} \label{baza}
|kl;mn\rangle := {\mathcal{A}}\Big\{\varphi_{a,k}(q_1)\,\varphi_{a,l}(q_2)\,\varphi_{b,m}(q_3)\,\varphi_{b,n}(q_4)\Big\},
\end{equation}
where $\varphi_{a,k}(q)$ are eigenstates of corresponding single-particle harmonic oscillators, i.e.,
\begin{equation}
{\cal H}_\lambda\,\varphi_{\lambda,k}(q) = \left(k+\frac{1}{2}\right)\hbar\omega\,\varphi_{\lambda,k}(q)
\end{equation}
and $\mathcal{A}\big\{.\big\}$ is the anti-symmetrization operator in the appropriate subspace of indistinguishable fermions assuring that:
\begin{equation} \label{rules}
|kl;mn\rangle = -|lk;mn\rangle=-|kl;nm\rangle.
\end{equation}
Assuming some sufficiently large cutoff $k\leq N_{max}$ of the considered single-particle excitations one can calculate all matrix elements of the Hamiltonian Eq.~\eqref{Ham}. The resulting matrix is  diagonalized to find the exact ground state of the system $|\Phi(g)\rangle$ and its energy $E(g)$. In our case, the exact diagonalization is performed with the Arnoldi method \cite{lehoucq1998arpack} that was used previously with a great success for similar models \cite{SowinskiGrass2013FewInteracting,SowinskiEEL2015,sowinski2015slightly,Pecak2016Separation,Pecak2016Transition}. Alternative diagonalization routines that exploit effective interactions are also very efficient for all interaction
strengths \cite{RotureauEPJD2013,LindgrenNJoP2014}, although these methods have yet to be extended to the case with particles of different mass.

In the following, many-body wave functions in position representation corresponding to states $|\Psi(g)\rangle$ and $|\Phi(g)\rangle$ will be denoted as $\Psi_g(q_1,q_2;q_3,q_4)$ and $\Phi_g(q_1,q_2;q_3,q_4)$, respectively. Additionally, we introduce a dimensionless parameter $\mu=m_a/m_b$ for the mass ratio of atoms from different components.

\section{Quality of the ansatz wave function}
\begin{figure}[t]
\centering
\includegraphics[width=\columnwidth]{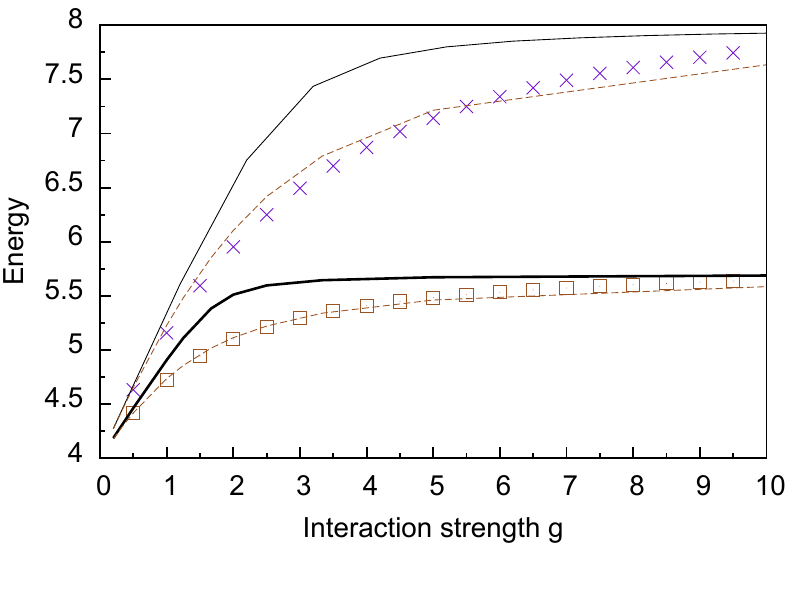}
\caption{Ground-state energy as a function of interactions predicted by the interpolatory ansatz Eq.~\eqref{variat} (solid thin and solid thick lines for $\mu=1$ and $\mu=10$, respectively) and numerically exact diagonalization of the Hamiltonian (crosses and squares for $\mu=1$ and $\mu=10$, respectively). Predictions of numerical ansatz are clearly overestimated and convergence to exact results is rather poor. Dashed lines correspond to modified ansatz, which gives much better predictions. See the main text for details. The energies and the interaction strength are measured in units of $\hbar \omega$ and $\hbar^{3/2} \omega^{1/2} m_b^{-1/2}$, respectively.}
\label{Fig1}
\end{figure}

{\bf The ground-state energy.}---
The quality of the assumed form of the variational wave function can be examined in various ways depending on the physical quantity one is interested in. Before any sophisticated tests are performed one should check predictions for the energy of the ground state since this quantity is always bounded from below by the exact value of the ground-state energy. Moreover, the energy of the ground state is a quantity, which in systems of few ultra-cold particles can be measured experimentally
with high accuracy \cite{zurnPRL2012,WenzS2013}.

To test the predictions of the variational method based on this natural quantity we compare the variational energy of the ground state with its counterpart obtained with the exact diagonalization method. The results are presented in Fig.~\ref{Fig1}, where solid lines represent variational ansatz predictions whereas crosses and squares correspond to the exact-diagonalization predictions (see caption of Fig.~\ref{Fig1} for details). Quite obviously, the energy is well reproduced in the limiting cases of $g=0$ and $g=\infty$. However, for the intermediate interactions the energy is clearly overestimated. Moreover, in the perturbation regime of small interactions ($g\approx 0$) the slope $\partial E(g)/\partial g|_{g=0}$ is not predicted correctly.
These results could suggest that the variational assumption that the ground state of the system can be well approximated with a simple superposition of two many-body eigenstates in limiting cases is maybe too simple.

At this point it is worth noticing, that the variational ansatz we use can be essentially improved to make predictions of the ground-state energy much more accurately. The modification is extensively described in \cite{AndersenAe2016} and briefly discussed in the Appendix. The improved results obtained in this framework are presented in Fig.~\ref{Fig1} by dashed lines. It is clear that the improvement of the resulting energies is essential. Nevertheless, as shown in \cite{AndersenAe2016}, in this case one loses accuracy in the predictions of the many-body wave functions. Therefore, in further discussion of other quantities the original ansatz Eq. \eqref{variat} is used and the modified ansatz is adopted only when displaying the energy spectrum in Fig.~\ref{Fig1}.

Let us note here that also the exact diagonalization method has some problems, mostly in the limit of very strong interactions. It is related to the fact that the resulting energies converge to the exact value very slowly with increasing cutoff $N_{max}$. Nevertheless, in principle one has a full control on this convergence and can unambiguously indicate a systematic error related to this numerical approximation. However, in cases where convergence is prohibitively slow, the access to a simple ansatz is extremely useful.

\begin{figure}[t]
\centering
\includegraphics[width=\columnwidth]{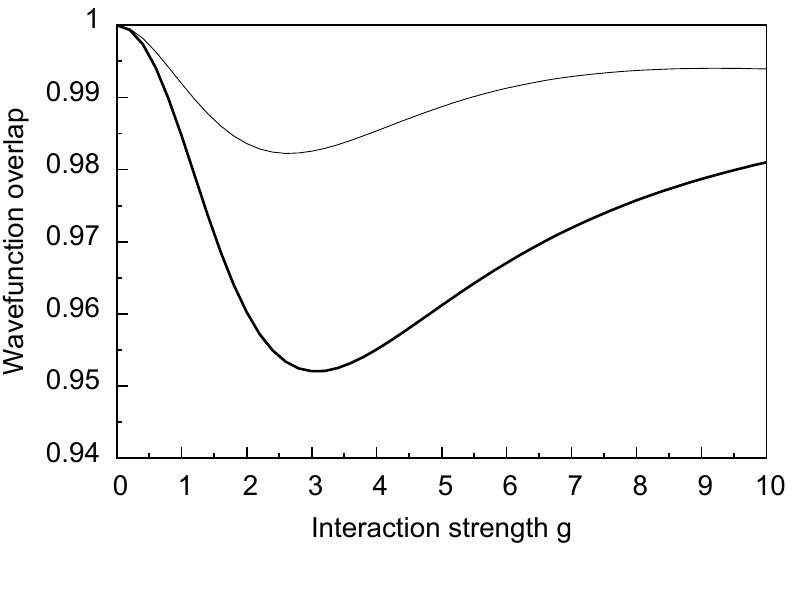}
\caption{The fidelity Eq.~\eqref{Fidel} between ground-state wave functions obtained variationally and with the exact diagonalization of the Hamiltonian (thin black and thick black line for system with $\mu=1$ and $\mu=10$, respectively). By the construction, in the limiting cases of vanishing or very strong interactions the fidelity is equal to $1$. For intermediate interactions, where the predictions of the ansatz are not exact, fidelity drops down. These results suggest that for system of different masses an inaccuracy is larger than for systems of the same mass. The variational ansatz results disaplayed here correspond to the original ansatz as discussed in the text. The interaction strength $g$ is measured in units of $\hbar^{3/2} \omega^{1/2} m_b^{-1/2}$.}
\label{Fig2}
\end{figure}

{\bf Overlap of the many-body ground states.}---
It is quite natural that in the case of any variational method used to determine the ground state of a many-body problem, a coincidence of energies is not sufficient to claim that the quantum state is predicted correctly. One of the methods to check if the quantum state is reproduced correctly is to calculate its fidelity, i.e., an overlap of the approximate state with the many-body ground state obtained from the exact diagonalization method:
\begin{equation} \label{Fidel}
{\cal F}(g) = |\langle \Psi(g)|\Phi(g)\rangle|^2.
\end{equation}
Obviously in the case studied, for $g=0$ and $g\rightarrow \infty$, the fidelity $\cal F$ is equal to 1 since in these limiting cases the wave function is reproduced exactly. For intermediate interactions the fidelity is smaller than $1$, and it is presented in Fig. 2. Surprisingly, for equal mass mixture $\mu=1$ (thin line) the overlap is close to $1$ for any interaction, i.e. the wave function of the ground state is reproduced correctly. However, if the mass ratio increases (thick line) the predictions of the ansatz become worse for intermediate interactions. However, the overlap is still quite large. This observation suggests that different quantities extracted from the approximate ground-state wave function served by the ansatz may have values close to those obtained from the exact method.

To check this hypothesis, in the following we will compare different predictions of the variational approximation with predictions of the exact diagonalization method.

\begin{figure}[t]
\centering
\includegraphics[width=\columnwidth]{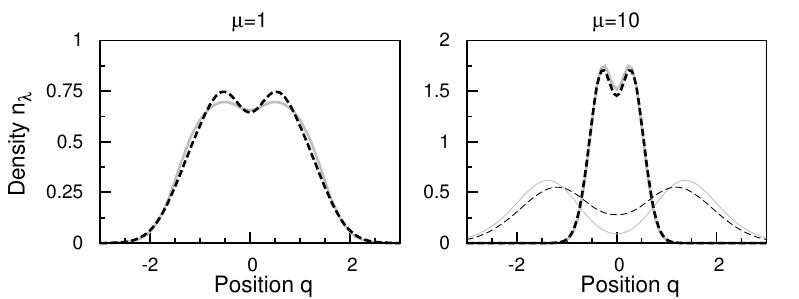}
\caption{Single-particle density profile for intermediate interactions, $g=2$, calculated from the exact ground state (solid lines) and from variational ground state (dashed lines). Density profiles are reproduced by the ansatz method quite well for equal-mass system $\mu=1$ (left panel) as well as for different mass systems $\mu=10$ (right panel). In the latter case the predictions are much better for heavier component (thick lines) than for lighter component (thin lines). Comparison of profiles for the lighter components suggests that the single-particle density undergoes a separation which is much sharper than that predicted by the variational method at the same interaction strength. As in Fig.~\ref{Fig2}, the ansatz results are based on the original ansatz. The positions and the densities are measured in units of $\sqrt{\hbar/(m_b \omega)}$ and $\sqrt{m_b \omega/\hbar}$, respectively. }
\label{Fig3}
\end{figure}
{\bf Single particle density.}---
Apart from the ground-state energy, one of the quantities, which can be measured straightforwardly in experiments, is a spatial density profile of the particles of a given component. Typically it is done by repeating and averaging instantaneous detections of positions of all particles. In principle, in the limit of an infinite number of repetitions, the resulting density approaches the theoretical quantities extracted from the many-body wave function
\begin{subequations}
\begin{align}
n_a(q_1) = \int\!\mathrm{d}q_2\int\!\mathrm{d}q_3\int\!\mathrm{d}q_4\,|\Phi_g(q_1,q_2;q_3,q_4)|^2, \\
n_b(q_3) = \int\!\mathrm{d}q_1\int\!\mathrm{d}q_2\int\!\mathrm{d}q_4\,|\Phi_g(q_1,q_2;q_3,q_4)|^2.
\end{align}
\end{subequations}
These profiles can be directly compared with the profiles calculated analogously from the variational ground state of the system $|\Psi(g)\rangle$. Obviously, since the ansatz is based on the proper wave functions in $g=0$ and $g\rightarrow\infty$, in these limiting cases the predictions of both methods match. If any discrepancies between both predictions exist, one should expect them in the range of interactions where the fidelity $\cal F$ is essentially less than 1. In Fig.~\ref{Fig3} we show the density profiles obtained from both methods for $g=2$. For equal mass case $\mu=1$ (left panel in Fig.~\ref{Fig3}), the exact profile is much flatter than the profile from the variational method. It means that for intermediate interactions variational wave function overestimate contribution from the non-interacting many-body wave function.

When the mass difference between atoms is introduced (right panel in Fig.~\ref{Fig3}), the density profile of a heavier component is improved. At the same time, the density of a lighter component becomes worse. We checked that this scenario is quite general and it does not depend on statistics, i.e., the result is the same when analogous variational method is adopted for Bose-Bose or Bose-Fermi mixtures.

Although the density profiles predicted by the variational ansatz have some discrepancies when compared to exact results, these differences are rather marginal and should not be of importance in comparison to experimental results. Additionally, we checked that also on the level of a complete single-particle density matrix (not only on its diagonal part) the predictions of the variational ansatz are very close to the exact results. It means that the proposed variational wave function can be safely used to predict any single-particle properties of the system of equal as well as different masses.

\begin{figure}[t]
\centering
\includegraphics[width=\columnwidth]{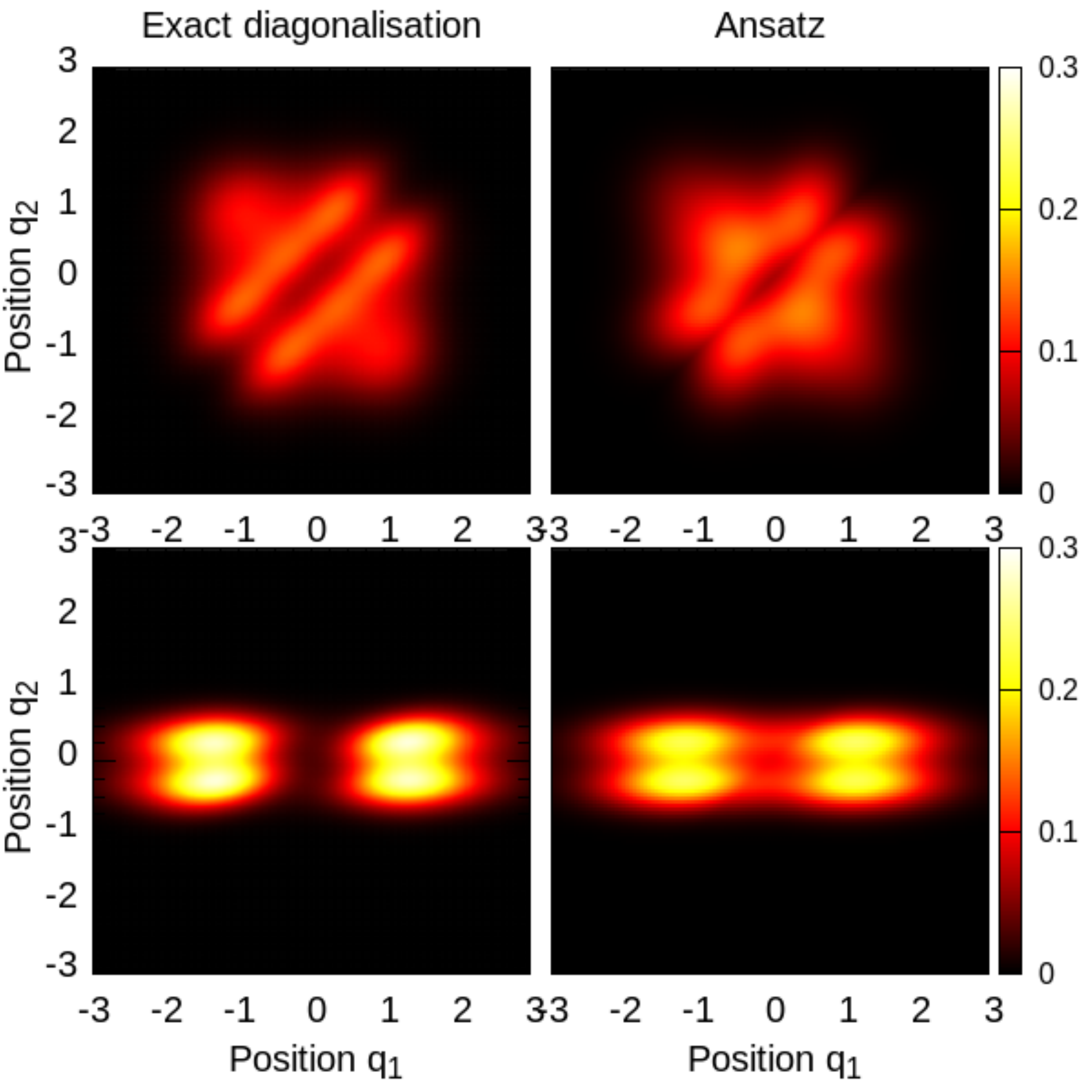}
\caption{Two-body density profile of opposite fermions calculated in the ground state of the system calculated with both methods for intermediate interaction $g=2$. Predictions of the variational method (right panels) are consistent with predictions of the exact diagonalization (left panels) for equal-mass $\mu=1$ (top panels) as well as different mass $\mu=10$ (bottom panels) systems. However, in the latter case, i.e. for the mass imbalance $\mu=10$, the probability of finding both fermions in the middle of the trap is overestimated. This fact has consequences in the single-particle profiles (right panel of Fig.~\ref{Fig3}) where incomplete separation of densities is predicted by variational method.
Again, the ansatz results are based on the orginal ansatz as in Fig.~\ref{Fig2} and Fig.~\ref{Fig3}.
The positions  $q_1, q_2$ are measured in natural units of harmonic oscillator $\sqrt{\hbar/(m_b \omega)}$, and the two-body density is measured in units of $m_b \omega/\hbar$.}
\label{Fig4}
\end{figure}

{\bf Interparticle correlations.}---
A natural question, which arises at this point, is related to different interparticle correlations that are beyond description of a single-particle density matrix. Since the ansatz is based on a very simple superposition of two many-body states, in principle, it is not obvious if mutual correlations between particles, which are very sensitive to any change in the many-body wave function, are restored correctly. To answer this question we concentrate on the simplest two-body correlation, i.e., a two-particle density profile between components defined as
\begin{align}
\rho(q_1,q_3) = \int\!\mathrm{d}q_2\int\!\mathrm{d}q_4\,|\Phi_g(q_1,q_2;q_3,q_4)|^2.
\end{align}

Density profiles for interacting system of four fermions of the same and different masses are presented in top and bottom panels of Fig.~\ref{Fig4}, respectively. As before, the presented results are obtained for the intermediate interactions $g=2$, where the fidelity $\cal F$ is essentially lower than 1. It is seen that, in general, the predictions of the variational method are also consistent with exact results. However, some differences are visible, especially for the different-mass systems. Firstly, the variational pair density profiles are much more smeared than the profiles obtained with the exact method. In addition, for the different-mass system, the exact probability of finding both particles in the middle of the trap, in contrast to predictions of the variational method, rapidly drops to zero when the mass ratio $\mu$ is increased. This observation is one of discrepancies of the variational ansatz, which may lead to some quantitative differences as compared to experimental data.

\begin{figure}[t]
\centering
\includegraphics[width=\columnwidth]{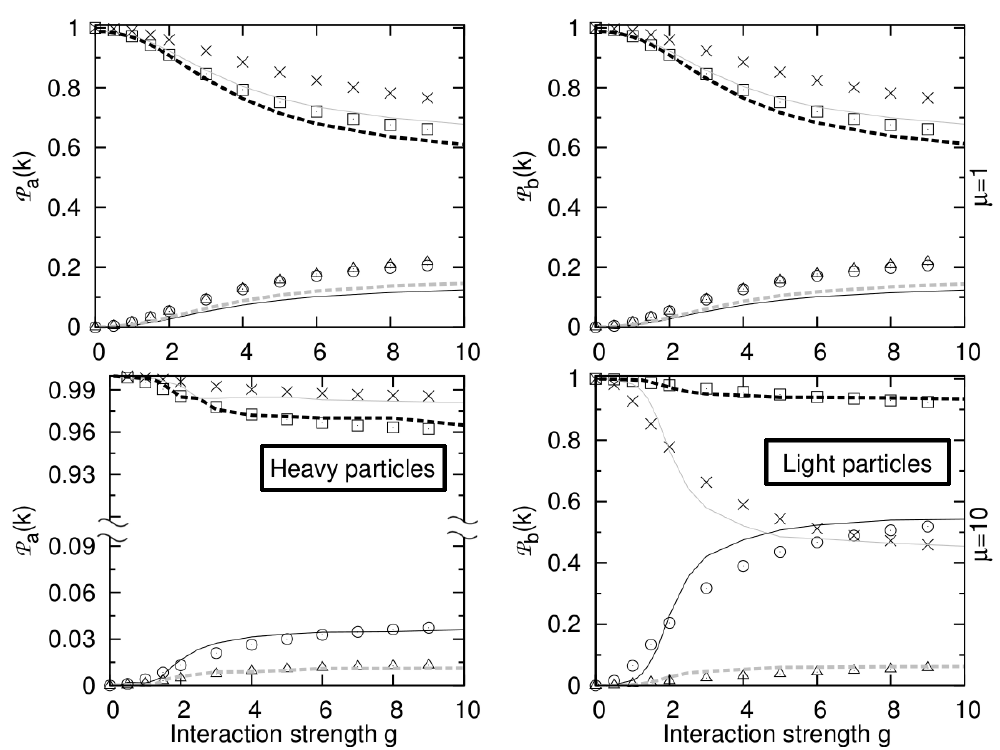}
\caption{The probabilities Eq.~\eqref{prob} of finding a single fermion in a given single-particle orbital of the harmonic confinement as functions of interactions. In the limit of vanishing interactions the fermions can be found only in the two the lowest orbitals. When the interactions are present other orbitals contribute to the ground state of the system. The predictions of the variational method (solid grey for $k=0$, dashed black for $k=1$, solid black for $k=2$, and dashed grey for $k=3$) are roughly consistent with the exact diagonalization results (crosses for $k=0$, squares for $k=1$, circles for $k=2$, and triangles for $k=3$). However, for the mass imbalanced systems and stronger interactions the variational method predicts too rapid drop of the ground-orbital contribution below the contribution of the second excited orbital. Here the results are also obtained with the original
ansatz as done in Fig.~\ref{Fig2}, Fig.~\ref{Fig3}, and Fig.~\ref{Fig4}. The interaction strength $g$ is measured in units of $\hbar^{3/2} \omega^{1/2} m_b^{-1/2}$.}
\label{Fig5}
\end{figure}

{\bf Occupations.}---
One of the less obvious ways of comparing results obtained with different methods is checking the predictions for the occupations of the single-particle orbitals, i.e., the quantities which mathematically are defined for the variational ground state of the system $|\Psi(g)\rangle$ as
\begin{subequations} \label{prob}
\begin{align}
{\cal P}_{a}(k) &= \sum_{lmn}\langle kl;mn|\Psi(g)\rangle, \\
{\cal P}_{b}(m) &= \sum_{kln}\langle kl;mn|\Psi(g)\rangle.
\end{align}
\end{subequations}
For the exact ground state of the system $|\Phi(g)\rangle$ the definitions are analogous. These quantities are quite interesting in the context of ultra-cold atoms since they can be measured experimentally by an appropriate lowering of the external
confinement \cite{MurmannPRL2015a}. Therefore, the theoretical predictions for these quantities can be validated.

In Fig.~\ref{Fig5} we present probabilities Eq.~\eqref{prob} calculated for some of the lowest single-particle states as the functions of interactions $g$ for equal (top panel) and different (bottom panel) mass systems. The results based on the variational method (lines) are compared with the probabilities obtained from the exact-diagonalization approach (squares, crosses, etc.). Obviously, in the case of an equal-mass system, both flavors have exactly the same probabilities. For vanishing interactions particles can be found only in the two lowest states (black solid lines and crosses or black dashed lines and squares for states with $k=0$ or $k=1$, respectively). As interaction increases, both probabilities decrease and higher single-particle states become partially occupied. In this case, predictions of the variational method, although not perfect, reproduce results from the exact method quite well.

The situation changes significantly, when a mixture of different masses is considered. In this case, predictions of both methods are roughly consistent only for the heavier component. For the lighter component, the occupation of the lowest single-particle orbital rapidly drops with the increase of the interactions and the third orbital becomes significantly occupied. Moreover, at some moment the occupation of the ground orbital becomes less probable than the occupation of the third state. This behavior of probabilities for lighter component is predicted by the variational ansatz. However, for small interactions a mentioned drop is too slow, whereas for stronger interactions (around $g=2$) it is too rapid. Nevertheless, differences between exact diagonalization predictions and variational approach are not essential. It means that also in the case of different-mass systems the variational ansatz can still be used for qualitative predictions when occupations of different single-particle orbitals are considered.

This extremely good agreement between the predictions of the variational ansatz and the exact-diagonalization approach in the case of the heavy component is related to the fact that in both limiting cases ($g=0$ and $g\rightarrow \infty$) the heavy particles are located in the middle of the trap. The situation is different for light particles, i.e., in the limit of strong repulsion the light particles are pushed out from the middle of the trap. This implies that for the light particles interactions have a stronger effect spatially. As a consequence, it is considerably more difficult to capture this effect by the ansatz constructed as a superposition of the limiting wave functions. This effect is directly reflected in the occupations of the single-particle orbitals.

\section{Final remarks}
{\bf Other statistics.}---
Although the results presented are related to the fermionic mixtures, to obtain a wider perspective on the problem of accuracy of the variational ansatz, it is worth considering different kinds of mixtures of four quantum particles. Both methods, i.e., the variational ansatz and the exact diagonalization approach, can be easily adopted for mixtures of two kinds of bosons or one kind of boson and one kind of fermion. Formally, the only difference has to be introduced in (anti)-symmetrization definitions in Eq.~\eqref{rules}. Of course, these changes may have (and typically do have) decisive consequences for the results obtained.

We have performed appropriate calculations for Bose-Bose and Bose-Fermi mixtures under the assumption that the bosons within a given flavor do not interact and the only non-vanishing interaction is present between different components. This assumption gives us a simple comprehensive tool for testing the role of quantum statistics. The strongly interacting states for the equal mass case in Bose-Fermi mixtures have been a subject of several recent discussions \cite{HuNJoP2016,deuretzbacher2016spin,Dehkharghani2017}. We note that in Bose-Bose mixtures with no interactions within a given flavor, the strongly interacting wave function cannot be found by building it on a basis of totally antisymmetric wave function, but other techniques to obtain it have been discussed recently \cite{DehkharghaniSR2015,DehkharghaniPRA2015,DehkharghaniJoPBAMaOP2016}.

While we do not present the full results of our calculations here, some of the results obtained in this way were already mentioned previously. From our numerical tests and comparisons some general conclusions about the role of the statistics can be given. Independently of the statistics, the simple variational ansatz works surprisingly well and it can be safely used for simple qualitative and quantitative predictions when single-particle observables are considered. In fact, the case that we have presented here with two kinds of fermions is that in which the comparison between the numerically exact and the variational method is the worst. For other compositions of the particles the variational method agrees even better with the numerical results. We do caution though that whenever higher interparticle correlations are considered, one should be very careful since the predictions of the variational method proposed can be overestimated.

{\bf Improving the ansatz.}---
It is quite obvious that in principle the variational probe function \eqref{variat} could be extended by superposing additional many-body state --- for example the many-body ground state obtained numerically for the interaction $g$ for which the accuracy is the worst. Although such extension is possible it requires some numerical effort to obtain an additional many-body state. Therefore, a lot of the beauty and simplicity of the idea may be quickly lost. Nevertheless, this direction would be necessary if larger number of particles were considered.

{\bf Conclusions.}---
In this paper we compared predictions of the interpolatory ansatz introduced in \cite{AndersenAe2016} with the numerically exact method of diagonalization of the many-body Hamiltonian. Surprisingly, the simple assumption that the ground state of four interacting fermions of different masses can be well approximated by a superposition of two many-body ground states obtained in the limits of very strong and vanishing interactions, is sufficient to describe many different properties of the system. Obviously, in this simplified description some discrepancies are present for the intermediate interactions, but they are rather small and not decisive in the view of a quite drastic simplification.

\section{Acknowledgments}
This work was partially supported by the (Polish) National Science Center Grants No. 2016/21/N/ST2/03315 (DP) and 2016/22/E/ST2/00555 (TS). The authors would like to thank M.~E.~S. Andersen, M. Valiente, and A. Volosniev for discussions. The work of A.~S. D. and N.~T. Z. is supported by the Danish Council for Independent Research DFF and the DFF Sapere Aude program.

\appendix

\section*{Appendix: Details of the ansatz}\label{details}
It can be shown straightforwardly that the trial energy $E$ calculated as an expectation value of the Hamiltonian \eqref{Ham} in the variational wave function \eqref{variat} is given as:
\begin{align*}
E = \frac{\langle{\Psi(g)|\mathcal{H}|\Psi(g)}\rangle}{\langle{\Psi(g)|\Psi(g)}\rangle} = E(0) + \frac{\langle{\Psi_0|{\cal H}_{ab}|\Psi_0}\rangle\alpha^2 + \Delta E\beta^2}{\alpha^2 + \beta^2 + 2\langle{\Psi_0|\Psi_\infty}\rangle\alpha\beta},
\end{align*}
where $\Delta E = E(\infty) - E(0)$. By finding the extreme points of the above expression, one finds the stationary solutions, which are determined by the following condition
\begin{widetext}
\begin{equation}
\left(\frac{\alpha}{\beta}\right)_\mathrm{opt}^{(\pm)} = \frac{\Delta E - \langle{\Psi_0|{\cal H}_{ab}|\Psi_0}\rangle
\mp\sqrt{\left(\Delta E - \langle{\Psi_0|{\cal H}_{ab}|\Psi_0}\rangle\right)^2+4 \langle{\Psi_0|{\cal H}_{ab}|\Psi_0}\rangle\Delta E \langle{\Psi_0|\Psi_\infty}\rangle^2}}{2 \langle{\Psi_0|{\cal H}_{ab}|\Psi_0}\rangle\langle {\Psi_0|\Psi_\infty}\rangle}.
\end{equation}
As a consequence, the optimized energy reduces to
\begin{equation} \label{OptEnergy}
E_\mathrm{opt}^{(\pm)} = E(0) +\frac{ \langle\Psi_0|{\cal H}_{ab}|\Psi_0\rangle + \Delta E
\pm \sqrt{( \langle{\Psi_0|{\cal H}_{ab}|\Psi_0}\rangle + \Delta E)^2 - 4 \langle{\Psi_0|{\cal H}_{ab}|\Psi_0}\rangle \Delta E\left(1- \langle{\Psi_0|\Psi_\infty}\rangle^2\right)}}{2\left(1- \langle{\Psi_0|\Psi_\infty}\rangle^2\right)}.
\end{equation}
\end{widetext}
The above estimation of the energy turns out to be insufficient in the limit of strong as well as weak interactions \cite{AndersenAe2016}. It can be improved by requiring the correct slope of the energy as a function of $-1/g$. It is shown that up to the first-order expansion the slope of the energy in the strong interactions regime can be calculated exactly and it is given as  \cite{VolosnievNC2014}:
\begin{equation}
K_\mathrm{opt}^\infty = \frac{\partial E_\mathrm{opt}}{\partial (-1/g)}\bigg|_{g\rightarrow\infty} = \frac{\Delta E^2}{K^0}  \langle{\Psi_0|\Psi_\infty}\rangle^2,
\end{equation}
where $K^0 =  \langle{\Psi_0|{\cal H}_{ab}|\Psi_0}\rangle/g$. For a given $K_\mathrm{opt}^\infty$ we can now find a new value of $\langle{\Psi_0|\Psi_\infty}\rangle$, which can be inserted in the expression for the optimized energy \eqref{OptEnergy}. In this way one obtains much better estimation of the ground-state energy. Even though the modified ansatz reproduces the energy much better than the original ansatz (see Fig. \ref{Fig1}), it comes with the cost that the ground-state wave function is no longer known. As a consequence, the modified ansatz is useful only for performing a better estimation of the energy.

\bibliographystyle{apsrev4-1}
\bibliography{referencias}

\begin{thebibliography}{56}%
\makeatletter
\providecommand \@ifxundefined [1]{%
 \@ifx{#1\undefined}
}%
\providecommand \@ifnum [1]{%
 \ifnum #1\expandafter \@firstoftwo
 \else \expandafter \@secondoftwo
 \fi
}%
\providecommand \@ifx [1]{%
 \ifx #1\expandafter \@firstoftwo
 \else \expandafter \@secondoftwo
 \fi
}%
\providecommand \natexlab [1]{#1}%
\providecommand \enquote  [1]{``#1''}%
\providecommand \bibnamefont  [1]{#1}%
\providecommand \bibfnamefont [1]{#1}%
\providecommand \citenamefont [1]{#1}%
\providecommand \href@noop [0]{\@secondoftwo}%
\providecommand \href [0]{\begingroup \@sanitize@url \@href}%
\providecommand \@href[1]{\@@startlink{#1}\@@href}%
\providecommand \@@href[1]{\endgroup#1\@@endlink}%
\providecommand \@sanitize@url [0]{\catcode `\\12\catcode `\$12\catcode
  `\&12\catcode `\#12\catcode `\^12\catcode `\_12\catcode `\%12\relax}%
\providecommand \@@startlink[1]{}%
\providecommand \@@endlink[0]{}%
\providecommand \url  [0]{\begingroup\@sanitize@url \@url }%
\providecommand \@url [1]{\endgroup\@href {#1}{\urlprefix }}%
\providecommand \urlprefix  [0]{URL }%
\providecommand \Eprint [0]{\href }%
\providecommand \doibase [0]{http://dx.doi.org/}%
\providecommand \selectlanguage [0]{\@gobble}%
\providecommand \bibinfo  [0]{\@secondoftwo}%
\providecommand \bibfield  [0]{\@secondoftwo}%
\providecommand \translation [1]{[#1]}%
\providecommand \BibitemOpen [0]{}%
\providecommand \bibitemStop [0]{}%
\providecommand \bibitemNoStop [0]{.\EOS\space}%
\providecommand \EOS [0]{\spacefactor3000\relax}%
\providecommand \BibitemShut  [1]{\csname bibitem#1\endcsname}%
\let\auto@bib@innerbib\@empty
\bibitem [{\citenamefont {Lieb}\ and\ \citenamefont
  {Mattis}(1966)}]{lieb1966mathematical}%
  \BibitemOpen
  \bibfield  {author} {\bibinfo {author} {\bibfnamefont {E.~H.}\ \bibnamefont
  {Lieb}}\ and\ \bibinfo {author} {\bibfnamefont {D.~C.}\ \bibnamefont
  {Mattis}},\ }\href@noop {} {\emph {\bibinfo {title} {Mathematical physics in
  one dimension: exactly soluble models of interacting particles}}}\ (\bibinfo
  {publisher} {Academic Press},\ \bibinfo {address} {London},\ \bibinfo {year}
  {1966})\BibitemShut {NoStop}%
\bibitem [{\citenamefont {Baxter}(1989)}]{baxter1989exactly}%
  \BibitemOpen
  \bibfield  {author} {\bibinfo {author} {\bibfnamefont {R.}~\bibnamefont
  {Baxter}},\ }\href@noop {} {\emph {\bibinfo {title} {Exactly Solved Models in
  Statistical Mechanics}}}\ (\bibinfo  {publisher} {Academic Press},\ \bibinfo
  {address} {London},\ \bibinfo {year} {1989})\BibitemShut {NoStop}%
\bibitem [{\citenamefont {Giamarchi}(2003)}]{Giamarchi2003}%
  \BibitemOpen
  \bibfield  {author} {\bibinfo {author} {\bibfnamefont {T.}~\bibnamefont
  {Giamarchi}},\ }\href {https://books.google.dk/books?id=GVeuKZLGMZ0C} {\emph
  {\bibinfo {title} {Quantum Physics in One Dimension}}},\ International Series
  of Monographs on Physics\ (\bibinfo  {publisher} {Clarendon Press},\ \bibinfo
  {year} {2003})\BibitemShut {NoStop}%
\bibitem [{\citenamefont {Plischke}\ and\ \citenamefont
  {Bergersen}(2006)}]{plischke2006equilibrium}%
  \BibitemOpen
  \bibfield  {author} {\bibinfo {author} {\bibfnamefont {M.}~\bibnamefont
  {Plischke}}\ and\ \bibinfo {author} {\bibfnamefont {B.}~\bibnamefont
  {Bergersen}},\ }\href {https://books.google.dk/books?id=KYu7igYEkhwC} {\emph
  {\bibinfo {title} {Equilibrium Statistical Physics}}}\ (\bibinfo  {publisher}
  {World Scientific},\ \bibinfo {year} {2006})\BibitemShut {NoStop}%
\bibitem [{\citenamefont {Grining}\ \emph
  {et~al.}(2015{\natexlab{a}})\citenamefont {Grining}, \citenamefont {Tomza},
  \citenamefont {Lesiuk}, \citenamefont {Przybytek}, \citenamefont {Musial},
  \citenamefont {Massignan}, \citenamefont {Lewenstein},\ and\ \citenamefont
  {Moszynski}}]{GriningNJoP2015}%
  \BibitemOpen
  \bibfield  {author} {\bibinfo {author} {\bibfnamefont {T.}~\bibnamefont
  {Grining}}, \bibinfo {author} {\bibfnamefont {M.}~\bibnamefont {Tomza}},
  \bibinfo {author} {\bibfnamefont {M.}~\bibnamefont {Lesiuk}}, \bibinfo
  {author} {\bibfnamefont {M.}~\bibnamefont {Przybytek}}, \bibinfo {author}
  {\bibfnamefont {M.}~\bibnamefont {Musial}}, \bibinfo {author} {\bibfnamefont
  {P.}~\bibnamefont {Massignan}}, \bibinfo {author} {\bibfnamefont
  {M.}~\bibnamefont {Lewenstein}}, \ and\ \bibinfo {author} {\bibfnamefont
  {R.}~\bibnamefont {Moszynski}},\ }\href
  {http://stacks.iop.org/1367-2630/17/i=11/a=115001} {\bibfield  {journal}
  {\bibinfo  {journal} {New Journal of Physics}\ }\textbf {\bibinfo {volume}
  {17}},\ \bibinfo {pages} {115001} (\bibinfo {year}
  {2015}{\natexlab{a}})}\BibitemShut {NoStop}%
\bibitem [{\citenamefont {{Decamp}}\ \emph {et~al.}(2016)\citenamefont
  {{Decamp}}, \citenamefont {{Armagnat}}, \citenamefont {{Fang}}, \citenamefont
  {{Albert}}, \citenamefont {{Minguzzi}},\ and\ \citenamefont
  {{Vignolo}}}]{DecampAe2016}%
  \BibitemOpen
  \bibfield  {author} {\bibinfo {author} {\bibfnamefont {J.}~\bibnamefont
  {{Decamp}}}, \bibinfo {author} {\bibfnamefont {P.}~\bibnamefont
  {{Armagnat}}}, \bibinfo {author} {\bibfnamefont {B.}~\bibnamefont {{Fang}}},
  \bibinfo {author} {\bibfnamefont {M.}~\bibnamefont {{Albert}}}, \bibinfo
  {author} {\bibfnamefont {A.}~\bibnamefont {{Minguzzi}}}, \ and\ \bibinfo
  {author} {\bibfnamefont {P.}~\bibnamefont {{Vignolo}}},\ }\href@noop {}
  {\bibfield  {journal} {\bibinfo  {journal} {New J. Phys.}\ }\textbf {\bibinfo
  {volume} {18}},\ \bibinfo {pages} {055011} (\bibinfo {year}
  {2016})}\BibitemShut {NoStop}%
\bibitem [{\citenamefont {{Volosniev}}\ \emph {et~al.}(2014)\citenamefont
  {{Volosniev}}, \citenamefont {{Fedorov}}, \citenamefont {{Jensen}},
  \citenamefont {{Valiente}},\ and\ \citenamefont
  {{Zinner}}}]{VolosnievNC2014}%
  \BibitemOpen
  \bibfield  {author} {\bibinfo {author} {\bibfnamefont {A.~G.}\ \bibnamefont
  {{Volosniev}}}, \bibinfo {author} {\bibfnamefont {D.~V.}\ \bibnamefont
  {{Fedorov}}}, \bibinfo {author} {\bibfnamefont {A.~S.}\ \bibnamefont
  {{Jensen}}}, \bibinfo {author} {\bibfnamefont {M.}~\bibnamefont
  {{Valiente}}}, \ and\ \bibinfo {author} {\bibfnamefont {N.~T.}\ \bibnamefont
  {{Zinner}}},\ }\href {\doibase 10.1038/ncomms6300} {\bibfield  {journal}
  {\bibinfo  {journal} {Nature Communications}\ }\textbf {\bibinfo {volume}
  {5}},\ \bibinfo {eid} {5300} (\bibinfo {year} {2014})},\ \Eprint
  {http://arxiv.org/abs/1306.4610} {arXiv:1306.4610 [cond-mat.quant-gas]}
  \BibitemShut {NoStop}%
\bibitem [{\citenamefont {Dobrzyniecki}\ and\ \citenamefont
  {Sowi{\'n}ski}(2016)}]{Dobrzyniecki2016DoubleWell}%
  \BibitemOpen
  \bibfield  {author} {\bibinfo {author} {\bibfnamefont {J.}~\bibnamefont
  {Dobrzyniecki}}\ and\ \bibinfo {author} {\bibfnamefont {T.}~\bibnamefont
  {Sowi{\'n}ski}},\ }\href@noop {} {\bibfield  {journal} {\bibinfo  {journal}
  {EPJ D}\ }\textbf {\bibinfo {volume} {70}},\ \bibinfo {pages} {83} (\bibinfo
  {year} {2016})}\BibitemShut {NoStop}%
\bibitem [{\citenamefont {Ko{\'s}cik}(2012)}]{Koscik2012}%
  \BibitemOpen
  \bibfield  {author} {\bibinfo {author} {\bibfnamefont {P.}~\bibnamefont
  {Ko{\'s}cik}},\ }\href@noop {} {\bibfield  {journal} {\bibinfo  {journal}
  {Eur. Phys. J. B}\ }\textbf {\bibinfo {volume} {85}},\ \bibinfo {pages} {173}
  (\bibinfo {year} {2012})}\BibitemShut {NoStop}%
\bibitem [{\citenamefont {Ko{\'s}cik}(2015)}]{Koscik2015}%
  \BibitemOpen
  \bibfield  {author} {\bibinfo {author} {\bibfnamefont {P.}~\bibnamefont
  {Ko{\'s}cik}},\ }\href@noop {} {\bibfield  {journal} {\bibinfo  {journal}
  {Physics Letters A}\ }\textbf {\bibinfo {volume} {379}},\ \bibinfo {pages}
  {293} (\bibinfo {year} {2015})}\BibitemShut {NoStop}%
\bibitem [{\citenamefont {Ko{\'s}cik}(2017)}]{Koscik2017}%
  \BibitemOpen
  \bibfield  {author} {\bibinfo {author} {\bibfnamefont {P.}~\bibnamefont
  {Ko{\'s}cik}},\ }\href@noop {} {\bibfield  {journal} {\bibinfo  {journal}
  {Few-Body Systems}\ }\textbf {\bibinfo {volume} {58}},\ \bibinfo {pages} {59}
  (\bibinfo {year} {2017})}\BibitemShut {NoStop}%
\bibitem [{\citenamefont {Deuretzbacher}\ \emph {et~al.}(2014)\citenamefont
  {Deuretzbacher}, \citenamefont {Becker}, \citenamefont {Bjerlin},
  \citenamefont {Reimann},\ and\ \citenamefont
  {Santos}}]{DeuretzbacherPRA2014}%
  \BibitemOpen
  \bibfield  {author} {\bibinfo {author} {\bibfnamefont {F.}~\bibnamefont
  {Deuretzbacher}}, \bibinfo {author} {\bibfnamefont {D.}~\bibnamefont
  {Becker}}, \bibinfo {author} {\bibfnamefont {J.}~\bibnamefont {Bjerlin}},
  \bibinfo {author} {\bibfnamefont {S.~M.}\ \bibnamefont {Reimann}}, \ and\
  \bibinfo {author} {\bibfnamefont {L.}~\bibnamefont {Santos}},\ }\href
  {\doibase 10.1103/PhysRevA.90.013611} {\bibfield  {journal} {\bibinfo
  {journal} {Phys. Rev. A}\ }\textbf {\bibinfo {volume} {90}},\ \bibinfo
  {pages} {013611} (\bibinfo {year} {2014})}\BibitemShut {NoStop}%
\bibitem [{\citenamefont {Yang}\ \emph {et~al.}(2015)\citenamefont {Yang},
  \citenamefont {Guan},\ and\ \citenamefont {Pu}}]{YangPRA2015}%
  \BibitemOpen
  \bibfield  {author} {\bibinfo {author} {\bibfnamefont {L.}~\bibnamefont
  {Yang}}, \bibinfo {author} {\bibfnamefont {L.}~\bibnamefont {Guan}}, \ and\
  \bibinfo {author} {\bibfnamefont {H.}~\bibnamefont {Pu}},\ }\href {\doibase
  10.1103/PhysRevA.91.043634} {\bibfield  {journal} {\bibinfo  {journal} {Phys.
  Rev. A}\ }\textbf {\bibinfo {volume} {91}},\ \bibinfo {pages} {043634}
  (\bibinfo {year} {2015})}\BibitemShut {NoStop}%
\bibitem [{\citenamefont {Yang}\ and\ \citenamefont {Cui}(2016)}]{YangPRA2016}%
  \BibitemOpen
  \bibfield  {author} {\bibinfo {author} {\bibfnamefont {L.}~\bibnamefont
  {Yang}}\ and\ \bibinfo {author} {\bibfnamefont {X.}~\bibnamefont {Cui}},\
  }\href {\doibase 10.1103/PhysRevA.93.013617} {\bibfield  {journal} {\bibinfo
  {journal} {Phys. Rev. A}\ }\textbf {\bibinfo {volume} {93}},\ \bibinfo
  {pages} {013617} (\bibinfo {year} {2016})}\BibitemShut {NoStop}%
\bibitem [{\citenamefont {Gharashi}\ and\ \citenamefont
  {Blume}(2013)}]{GharashiPRL2013}%
  \BibitemOpen
  \bibfield  {author} {\bibinfo {author} {\bibfnamefont {S.~E.}\ \bibnamefont
  {Gharashi}}\ and\ \bibinfo {author} {\bibfnamefont {D.}~\bibnamefont
  {Blume}},\ }\href {\doibase 10.1103/PhysRevLett.111.045302} {\bibfield
  {journal} {\bibinfo  {journal} {Phys. Rev. Lett.}\ }\textbf {\bibinfo
  {volume} {111}},\ \bibinfo {pages} {045302} (\bibinfo {year}
  {2013})}\BibitemShut {NoStop}%
\bibitem [{\citenamefont {Garcia-March}\ \emph {et~al.}(2014)\citenamefont
  {Garcia-March}, \citenamefont {Julia-Diaz}, \citenamefont {Astrakharchik},
  \citenamefont {Busch}, \citenamefont {Boronat},\ and\ \citenamefont
  {Polls}}]{Garcia-MarchNJoP2014}%
  \BibitemOpen
  \bibfield  {author} {\bibinfo {author} {\bibfnamefont {M.~A.}\ \bibnamefont
  {Garcia-March}}, \bibinfo {author} {\bibfnamefont {B.}~\bibnamefont
  {Julia-Diaz}}, \bibinfo {author} {\bibfnamefont {G.~E.}\ \bibnamefont
  {Astrakharchik}}, \bibinfo {author} {\bibfnamefont {T.}~\bibnamefont
  {Busch}}, \bibinfo {author} {\bibfnamefont {J.}~\bibnamefont {Boronat}}, \
  and\ \bibinfo {author} {\bibfnamefont {A.}~\bibnamefont {Polls}},\ }\href
  {http://stacks.iop.org/1367-2630/16/i=10/a=103004} {\bibfield  {journal}
  {\bibinfo  {journal} {New Journal of Physics}\ }\textbf {\bibinfo {volume}
  {16}},\ \bibinfo {pages} {103004} (\bibinfo {year} {2014})}\BibitemShut
  {NoStop}%
\bibitem [{\citenamefont {Bloch}\ \emph {et~al.}(2008)\citenamefont {Bloch},
  \citenamefont {Dalibard},\ and\ \citenamefont {Zwerger}}]{blochRMP2008}%
  \BibitemOpen
  \bibfield  {author} {\bibinfo {author} {\bibfnamefont {I.}~\bibnamefont
  {Bloch}}, \bibinfo {author} {\bibfnamefont {J.}~\bibnamefont {Dalibard}}, \
  and\ \bibinfo {author} {\bibfnamefont {W.}~\bibnamefont {Zwerger}},\ }\href
  {\doibase 10.1103/RevModPhys.80.885} {\bibfield  {journal} {\bibinfo
  {journal} {Rev. Mod. Phys.}\ }\textbf {\bibinfo {volume} {80}},\ \bibinfo
  {pages} {885} (\bibinfo {year} {2008})}\BibitemShut {NoStop}%
\bibitem [{\citenamefont {Lewenstein}\ \emph {et~al.}(2007)\citenamefont
  {Lewenstein}, \citenamefont {Sanpera}, \citenamefont {Ahufinger},
  \citenamefont {Damski}, \citenamefont {Sen(De)},\ and\ \citenamefont
  {Sen}}]{lewensteinAiP2007}%
  \BibitemOpen
  \bibfield  {author} {\bibinfo {author} {\bibfnamefont {M.}~\bibnamefont
  {Lewenstein}}, \bibinfo {author} {\bibfnamefont {A.}~\bibnamefont {Sanpera}},
  \bibinfo {author} {\bibfnamefont {V.}~\bibnamefont {Ahufinger}}, \bibinfo
  {author} {\bibfnamefont {B.}~\bibnamefont {Damski}}, \bibinfo {author}
  {\bibfnamefont {A.}~\bibnamefont {Sen(De)}}, \ and\ \bibinfo {author}
  {\bibfnamefont {U.}~\bibnamefont {Sen}},\ }\href {\doibase
  10.1080/00018730701223200} {\bibfield  {journal} {\bibinfo  {journal}
  {Advances in Physics}\ }\textbf {\bibinfo {volume} {56}},\ \bibinfo {pages}
  {243} (\bibinfo {year} {2007})},\ \Eprint
  {http://arxiv.org/abs/http://dx.doi.org/10.1080/00018730701223200}
  {http://dx.doi.org/10.1080/00018730701223200} \BibitemShut {NoStop}%
\bibitem [{\citenamefont {Esslinger}(2010)}]{esslingerARoCMP2010}%
  \BibitemOpen
  \bibfield  {author} {\bibinfo {author} {\bibfnamefont {T.}~\bibnamefont
  {Esslinger}},\ }\href {\doibase 10.1146/annurev-conmatphys-070909-104059}
  {\bibfield  {journal} {\bibinfo  {journal} {Annual Review of Condensed Matter
  Physics}\ }\textbf {\bibinfo {volume} {1}},\ \bibinfo {pages} {129} (\bibinfo
  {year} {2010})},\ \Eprint
  {http://arxiv.org/abs/http://dx.doi.org/10.1146/annurev-conmatphys-070909-104059}
  {http://dx.doi.org/10.1146/annurev-conmatphys-070909-104059} \BibitemShut
  {NoStop}%
\bibitem [{\citenamefont {Moritz}\ \emph {et~al.}(2003)\citenamefont {Moritz},
  \citenamefont {St\"oferle}, \citenamefont {K\"ohl},\ and\ \citenamefont
  {Esslinger}}]{moritzPRL2003}%
  \BibitemOpen
  \bibfield  {author} {\bibinfo {author} {\bibfnamefont {H.}~\bibnamefont
  {Moritz}}, \bibinfo {author} {\bibfnamefont {T.}~\bibnamefont {St\"oferle}},
  \bibinfo {author} {\bibfnamefont {M.}~\bibnamefont {K\"ohl}}, \ and\ \bibinfo
  {author} {\bibfnamefont {T.}~\bibnamefont {Esslinger}},\ }\href {\doibase
  10.1103/PhysRevLett.91.250402} {\bibfield  {journal} {\bibinfo  {journal}
  {Phys. Rev. Lett.}\ }\textbf {\bibinfo {volume} {91}},\ \bibinfo {pages}
  {250402} (\bibinfo {year} {2003})}\BibitemShut {NoStop}%
\bibitem [{\citenamefont {St\"oferle}\ \emph {et~al.}(2004)\citenamefont
  {St\"oferle}, \citenamefont {Moritz}, \citenamefont {Schori}, \citenamefont
  {K\"ohl},\ and\ \citenamefont {Esslinger}}]{stoferlePRL2003}%
  \BibitemOpen
  \bibfield  {author} {\bibinfo {author} {\bibfnamefont {T.}~\bibnamefont
  {St\"oferle}}, \bibinfo {author} {\bibfnamefont {H.}~\bibnamefont {Moritz}},
  \bibinfo {author} {\bibfnamefont {C.}~\bibnamefont {Schori}}, \bibinfo
  {author} {\bibfnamefont {M.}~\bibnamefont {K\"ohl}}, \ and\ \bibinfo {author}
  {\bibfnamefont {T.}~\bibnamefont {Esslinger}},\ }\href {\doibase
  10.1103/PhysRevLett.92.130403} {\bibfield  {journal} {\bibinfo  {journal}
  {Phys. Rev. Lett.}\ }\textbf {\bibinfo {volume} {92}},\ \bibinfo {pages}
  {130403} (\bibinfo {year} {2004})}\BibitemShut {NoStop}%
\bibitem [{\citenamefont {Kinoshita}\ \emph {et~al.}(2004)\citenamefont
  {Kinoshita}, \citenamefont {Wenger},\ and\ \citenamefont
  {Weiss}}]{KinoshitaS2004}%
  \BibitemOpen
  \bibfield  {author} {\bibinfo {author} {\bibfnamefont {T.}~\bibnamefont
  {Kinoshita}}, \bibinfo {author} {\bibfnamefont {T.}~\bibnamefont {Wenger}}, \
  and\ \bibinfo {author} {\bibfnamefont {D.~S.}\ \bibnamefont {Weiss}},\ }\href
  {\doibase 10.1126/science.1100700} {\bibfield  {journal} {\bibinfo  {journal}
  {Science}\ }\textbf {\bibinfo {volume} {305}},\ \bibinfo {pages} {1125}
  (\bibinfo {year} {2004})},\ \Eprint
  {http://arxiv.org/abs/http://science.sciencemag.org/content/305/5687/1125.full.pdf}
  {http://science.sciencemag.org/content/305/5687/1125.full.pdf} \BibitemShut
  {NoStop}%
\bibitem [{\citenamefont {Kinoshita}\ \emph {et~al.}(2006)\citenamefont
  {Kinoshita}, \citenamefont {Wenger},\ and\ \citenamefont
  {Weiss}}]{KinoshitaN2006}%
  \BibitemOpen
  \bibfield  {author} {\bibinfo {author} {\bibfnamefont {T.}~\bibnamefont
  {Kinoshita}}, \bibinfo {author} {\bibfnamefont {T.}~\bibnamefont {Wenger}}, \
  and\ \bibinfo {author} {\bibfnamefont {D.~S.}\ \bibnamefont {Weiss}},\ }\href
  {\doibase 10.1038/nature04693} {\bibfield  {journal} {\bibinfo  {journal}
  {Nature}\ }\textbf {\bibinfo {volume} {440}},\ \bibinfo {pages} {900}
  (\bibinfo {year} {2006})}\BibitemShut {NoStop}%
\bibitem [{\citenamefont {Paredes}\ \emph {et~al.}(2004)\citenamefont
  {Paredes}, \citenamefont {Widera}, \citenamefont {Murg}, \citenamefont
  {Mandel}, \citenamefont {Folling}, \citenamefont {Cirac}, \citenamefont
  {Shlyapnikov}, \citenamefont {Hansch},\ and\ \citenamefont
  {Bloch}}]{ParedesN2004}%
  \BibitemOpen
  \bibfield  {author} {\bibinfo {author} {\bibfnamefont {B.}~\bibnamefont
  {Paredes}}, \bibinfo {author} {\bibfnamefont {A.}~\bibnamefont {Widera}},
  \bibinfo {author} {\bibfnamefont {V.}~\bibnamefont {Murg}}, \bibinfo {author}
  {\bibfnamefont {O.}~\bibnamefont {Mandel}}, \bibinfo {author} {\bibfnamefont
  {S.}~\bibnamefont {Folling}}, \bibinfo {author} {\bibfnamefont
  {I.}~\bibnamefont {Cirac}}, \bibinfo {author} {\bibfnamefont {G.~V.}\
  \bibnamefont {Shlyapnikov}}, \bibinfo {author} {\bibfnamefont {T.~W.}\
  \bibnamefont {Hansch}}, \ and\ \bibinfo {author} {\bibfnamefont
  {I.}~\bibnamefont {Bloch}},\ }\href {\doibase 10.1038/nature02530} {\bibfield
   {journal} {\bibinfo  {journal} {Nature}\ }\textbf {\bibinfo {volume}
  {429}},\ \bibinfo {pages} {277} (\bibinfo {year} {2004})}\BibitemShut
  {NoStop}%
\bibitem [{\citenamefont {Haller}\ \emph {et~al.}(2009)\citenamefont {Haller},
  \citenamefont {Gustavsson}, \citenamefont {Mark}, \citenamefont {Danzl},
  \citenamefont {Hart}, \citenamefont {Pupillo},\ and\ \citenamefont
  {N{\"a}gerl}}]{HallerS2009}%
  \BibitemOpen
  \bibfield  {author} {\bibinfo {author} {\bibfnamefont {E.}~\bibnamefont
  {Haller}}, \bibinfo {author} {\bibfnamefont {M.}~\bibnamefont {Gustavsson}},
  \bibinfo {author} {\bibfnamefont {M.~J.}\ \bibnamefont {Mark}}, \bibinfo
  {author} {\bibfnamefont {J.~G.}\ \bibnamefont {Danzl}}, \bibinfo {author}
  {\bibfnamefont {R.}~\bibnamefont {Hart}}, \bibinfo {author} {\bibfnamefont
  {G.}~\bibnamefont {Pupillo}}, \ and\ \bibinfo {author} {\bibfnamefont
  {H.-C.}\ \bibnamefont {N{\"a}gerl}},\ }\href {\doibase
  10.1126/science.1175850} {\bibfield  {journal} {\bibinfo  {journal}
  {Science}\ }\textbf {\bibinfo {volume} {325}},\ \bibinfo {pages} {1224}
  (\bibinfo {year} {2009})},\ \Eprint
  {http://arxiv.org/abs/http://science.sciencemag.org/content/325/5945/1224.full.pdf}
  {http://science.sciencemag.org/content/325/5945/1224.full.pdf} \BibitemShut
  {NoStop}%
\bibitem [{\citenamefont {Haller}\ \emph {et~al.}(2010)\citenamefont {Haller},
  \citenamefont {Hart}, \citenamefont {Mark}, \citenamefont {Danzl},
  \citenamefont {Reichs{\"o}llner}, \citenamefont {Gustavsson}, \citenamefont
  {Dalmonte}, \citenamefont {Pupillo},\ and\ \citenamefont
  {N{\"a}gerl}}]{haller2010pinning}%
  \BibitemOpen
  \bibfield  {author} {\bibinfo {author} {\bibfnamefont {E.}~\bibnamefont
  {Haller}}, \bibinfo {author} {\bibfnamefont {R.}~\bibnamefont {Hart}},
  \bibinfo {author} {\bibfnamefont {M.~J.}\ \bibnamefont {Mark}}, \bibinfo
  {author} {\bibfnamefont {J.~G.}\ \bibnamefont {Danzl}}, \bibinfo {author}
  {\bibfnamefont {L.}~\bibnamefont {Reichs{\"o}llner}}, \bibinfo {author}
  {\bibfnamefont {M.}~\bibnamefont {Gustavsson}}, \bibinfo {author}
  {\bibfnamefont {M.}~\bibnamefont {Dalmonte}}, \bibinfo {author}
  {\bibfnamefont {G.}~\bibnamefont {Pupillo}}, \ and\ \bibinfo {author}
  {\bibfnamefont {H.-C.}\ \bibnamefont {N{\"a}gerl}},\ }\href@noop {}
  {\bibfield  {journal} {\bibinfo  {journal} {Nature}\ }\textbf {\bibinfo
  {volume} {466}},\ \bibinfo {pages} {597} (\bibinfo {year}
  {2010})}\BibitemShut {NoStop}%
\bibitem [{\citenamefont {Pagano}\ \emph {et~al.}(2014)\citenamefont {Pagano},
  \citenamefont {Mancini}, \citenamefont {Cappellini}, \citenamefont
  {Lombardi}, \citenamefont {Schafer}, \citenamefont {Hu}, \citenamefont {Liu},
  \citenamefont {Catani}, \citenamefont {Sias}, \citenamefont {Inguscio},\ and\
  \citenamefont {Fallani}}]{PaganoNP2014}%
  \BibitemOpen
  \bibfield  {author} {\bibinfo {author} {\bibfnamefont {G.}~\bibnamefont
  {Pagano}}, \bibinfo {author} {\bibfnamefont {M.}~\bibnamefont {Mancini}},
  \bibinfo {author} {\bibfnamefont {G.}~\bibnamefont {Cappellini}}, \bibinfo
  {author} {\bibfnamefont {P.}~\bibnamefont {Lombardi}}, \bibinfo {author}
  {\bibfnamefont {F.}~\bibnamefont {Schafer}}, \bibinfo {author} {\bibfnamefont
  {H.}~\bibnamefont {Hu}}, \bibinfo {author} {\bibfnamefont {X.-J.}\
  \bibnamefont {Liu}}, \bibinfo {author} {\bibfnamefont {J.}~\bibnamefont
  {Catani}}, \bibinfo {author} {\bibfnamefont {C.}~\bibnamefont {Sias}},
  \bibinfo {author} {\bibfnamefont {M.}~\bibnamefont {Inguscio}}, \ and\
  \bibinfo {author} {\bibfnamefont {L.}~\bibnamefont {Fallani}},\ }\href
  {http://dx.doi.org/10.1038/nphys2878} {\bibfield  {journal} {\bibinfo
  {journal} {Nat Phys}\ }\textbf {\bibinfo {volume} {10}},\ \bibinfo {pages}
  {198} (\bibinfo {year} {2014})},\ \bibinfo {note} {letter}\BibitemShut
  {NoStop}%
\bibitem [{\citenamefont {Murmann}\ \emph
  {et~al.}(2015{\natexlab{a}})\citenamefont {Murmann}, \citenamefont
  {Deuretzbacher}, \citenamefont {Z\"urn}, \citenamefont {Bjerlin},
  \citenamefont {Reimann}, \citenamefont {Santos}, \citenamefont {Lompe},\ and\
  \citenamefont {Jochim}}]{MurmannPRL2015a}%
  \BibitemOpen
  \bibfield  {author} {\bibinfo {author} {\bibfnamefont {S.}~\bibnamefont
  {Murmann}}, \bibinfo {author} {\bibfnamefont {F.}~\bibnamefont
  {Deuretzbacher}}, \bibinfo {author} {\bibfnamefont {G.}~\bibnamefont
  {Z\"urn}}, \bibinfo {author} {\bibfnamefont {J.}~\bibnamefont {Bjerlin}},
  \bibinfo {author} {\bibfnamefont {S.~M.}\ \bibnamefont {Reimann}}, \bibinfo
  {author} {\bibfnamefont {L.}~\bibnamefont {Santos}}, \bibinfo {author}
  {\bibfnamefont {T.}~\bibnamefont {Lompe}}, \ and\ \bibinfo {author}
  {\bibfnamefont {S.}~\bibnamefont {Jochim}},\ }\href {\doibase
  10.1103/PhysRevLett.115.215301} {\bibfield  {journal} {\bibinfo  {journal}
  {Phys. Rev. Lett.}\ }\textbf {\bibinfo {volume} {115}},\ \bibinfo {pages}
  {215301} (\bibinfo {year} {2015}{\natexlab{a}})}\BibitemShut {NoStop}%
\bibitem [{\citenamefont {Murmann}\ \emph
  {et~al.}(2015{\natexlab{b}})\citenamefont {Murmann}, \citenamefont
  {Bergschneider}, \citenamefont {Klinkhamer}, \citenamefont {Z\"urn},
  \citenamefont {Lompe},\ and\ \citenamefont {Jochim}}]{MurmannPRL2015b}%
  \BibitemOpen
  \bibfield  {author} {\bibinfo {author} {\bibfnamefont {S.}~\bibnamefont
  {Murmann}}, \bibinfo {author} {\bibfnamefont {A.}~\bibnamefont
  {Bergschneider}}, \bibinfo {author} {\bibfnamefont {V.~M.}\ \bibnamefont
  {Klinkhamer}}, \bibinfo {author} {\bibfnamefont {G.}~\bibnamefont {Z\"urn}},
  \bibinfo {author} {\bibfnamefont {T.}~\bibnamefont {Lompe}}, \ and\ \bibinfo
  {author} {\bibfnamefont {S.}~\bibnamefont {Jochim}},\ }\href {\doibase
  10.1103/PhysRevLett.114.080402} {\bibfield  {journal} {\bibinfo  {journal}
  {Phys. Rev. Lett.}\ }\textbf {\bibinfo {volume} {114}},\ \bibinfo {pages}
  {080402} (\bibinfo {year} {2015}{\natexlab{b}})}\BibitemShut {NoStop}%
\bibitem [{\citenamefont {Serwane}\ \emph {et~al.}(2011)\citenamefont
  {Serwane}, \citenamefont {Z{\"u}rn}, \citenamefont {Lompe}, \citenamefont
  {Ottenstein}, \citenamefont {Wenz},\ and\ \citenamefont
  {Jochim}}]{SerwaneS2011}%
  \BibitemOpen
  \bibfield  {author} {\bibinfo {author} {\bibfnamefont {F.}~\bibnamefont
  {Serwane}}, \bibinfo {author} {\bibfnamefont {G.}~\bibnamefont {Z{\"u}rn}},
  \bibinfo {author} {\bibfnamefont {T.}~\bibnamefont {Lompe}}, \bibinfo
  {author} {\bibfnamefont {T.~B.}\ \bibnamefont {Ottenstein}}, \bibinfo
  {author} {\bibfnamefont {A.~N.}\ \bibnamefont {Wenz}}, \ and\ \bibinfo
  {author} {\bibfnamefont {S.}~\bibnamefont {Jochim}},\ }\href {\doibase
  10.1126/science.1201351} {\bibfield  {journal} {\bibinfo  {journal}
  {Science}\ }\textbf {\bibinfo {volume} {332}},\ \bibinfo {pages} {336}
  (\bibinfo {year} {2011})},\ \Eprint
  {http://arxiv.org/abs/http://science.sciencemag.org/content/332/6027/336.full.pdf}
  {http://science.sciencemag.org/content/332/6027/336.full.pdf} \BibitemShut
  {NoStop}%
\bibitem [{\citenamefont {Wenz}\ \emph {et~al.}(2013)\citenamefont {Wenz},
  \citenamefont {Z{\"u}rn}, \citenamefont {Murmann}, \citenamefont {Brouzos},
  \citenamefont {Lompe},\ and\ \citenamefont {Jochim}}]{WenzS2013}%
  \BibitemOpen
  \bibfield  {author} {\bibinfo {author} {\bibfnamefont {A.~N.}\ \bibnamefont
  {Wenz}}, \bibinfo {author} {\bibfnamefont {G.}~\bibnamefont {Z{\"u}rn}},
  \bibinfo {author} {\bibfnamefont {S.}~\bibnamefont {Murmann}}, \bibinfo
  {author} {\bibfnamefont {I.}~\bibnamefont {Brouzos}}, \bibinfo {author}
  {\bibfnamefont {T.}~\bibnamefont {Lompe}}, \ and\ \bibinfo {author}
  {\bibfnamefont {S.}~\bibnamefont {Jochim}},\ }\href {\doibase
  10.1126/science.1240516} {\bibfield  {journal} {\bibinfo  {journal}
  {Science}\ }\textbf {\bibinfo {volume} {342}},\ \bibinfo {pages} {457}
  (\bibinfo {year} {2013})},\ \Eprint
  {http://arxiv.org/abs/http://science.sciencemag.org/content/342/6157/457.full.pdf}
  {http://science.sciencemag.org/content/342/6157/457.full.pdf} \BibitemShut
  {NoStop}%
\bibitem [{\citenamefont {Z\"urn}\ \emph {et~al.}(2013)\citenamefont {Z\"urn},
  \citenamefont {Wenz}, \citenamefont {Murmann}, \citenamefont {Bergschneider},
  \citenamefont {Lompe},\ and\ \citenamefont {Jochim}}]{zurnPRL2013}%
  \BibitemOpen
  \bibfield  {author} {\bibinfo {author} {\bibfnamefont {G.}~\bibnamefont
  {Z\"urn}}, \bibinfo {author} {\bibfnamefont {A.~N.}\ \bibnamefont {Wenz}},
  \bibinfo {author} {\bibfnamefont {S.}~\bibnamefont {Murmann}}, \bibinfo
  {author} {\bibfnamefont {A.}~\bibnamefont {Bergschneider}}, \bibinfo {author}
  {\bibfnamefont {T.}~\bibnamefont {Lompe}}, \ and\ \bibinfo {author}
  {\bibfnamefont {S.}~\bibnamefont {Jochim}},\ }\href {\doibase
  10.1103/PhysRevLett.111.175302} {\bibfield  {journal} {\bibinfo  {journal}
  {Phys. Rev. Lett.}\ }\textbf {\bibinfo {volume} {111}},\ \bibinfo {pages}
  {175302} (\bibinfo {year} {2013})}\BibitemShut {NoStop}%
\bibitem [{\citenamefont {{Z{\"u}rn}}\ \emph {et~al.}(2012)\citenamefont
  {{Z{\"u}rn}}, \citenamefont {{Serwane}}, \citenamefont {{Lompe}},
  \citenamefont {{Wenz}}, \citenamefont {{Ries}}, \citenamefont {{Bohn}},\ and\
  \citenamefont {{Jochim}}}]{zurnPRL2012}%
  \BibitemOpen
  \bibfield  {author} {\bibinfo {author} {\bibfnamefont {G.}~\bibnamefont
  {{Z{\"u}rn}}}, \bibinfo {author} {\bibfnamefont {F.}~\bibnamefont
  {{Serwane}}}, \bibinfo {author} {\bibfnamefont {T.}~\bibnamefont {{Lompe}}},
  \bibinfo {author} {\bibfnamefont {A.~N.}\ \bibnamefont {{Wenz}}}, \bibinfo
  {author} {\bibfnamefont {M.~G.}\ \bibnamefont {{Ries}}}, \bibinfo {author}
  {\bibfnamefont {J.~E.}\ \bibnamefont {{Bohn}}}, \ and\ \bibinfo {author}
  {\bibfnamefont {S.}~\bibnamefont {{Jochim}}},\ }\href {\doibase
  10.1103/PhysRevLett.108.075303} {\bibfield  {journal} {\bibinfo  {journal}
  {Physical Review Letters}\ }\textbf {\bibinfo {volume} {108}},\ \bibinfo
  {eid} {075303} (\bibinfo {year} {2012})},\ \Eprint
  {http://arxiv.org/abs/1111.2727} {arXiv:1111.2727 [cond-mat.quant-gas]}
  \BibitemShut {NoStop}%
\bibitem [{\citenamefont {Kaufman}\ \emph {et~al.}(2015)\citenamefont
  {Kaufman}, \citenamefont {Lester}, \citenamefont {{Foss-Feig}}, \citenamefont
  {Wall}, \citenamefont {Rey},\ and\ \citenamefont
  {Regal}}]{Kaufman2015Entangling}%
  \BibitemOpen
  \bibfield  {author} {\bibinfo {author} {\bibfnamefont {A.~M.}\ \bibnamefont
  {Kaufman}}, \bibinfo {author} {\bibfnamefont {B.~J.}\ \bibnamefont {Lester}},
  \bibinfo {author} {\bibfnamefont {M.}~\bibnamefont {{Foss-Feig}}}, \bibinfo
  {author} {\bibfnamefont {M.~L.}\ \bibnamefont {Wall}}, \bibinfo {author}
  {\bibfnamefont {A.~M.}\ \bibnamefont {Rey}}, \ and\ \bibinfo {author}
  {\bibfnamefont {C.}~\bibnamefont {Regal}},\ }\href@noop {} {\bibfield
  {journal} {\bibinfo  {journal} {Nature}\ }\textbf {\bibinfo {volume} {527}},\
  \bibinfo {pages} {208} (\bibinfo {year} {2015})}\BibitemShut {NoStop}%
\bibitem [{\citenamefont {Grining}\ \emph
  {et~al.}(2015{\natexlab{b}})\citenamefont {Grining}, \citenamefont {Tomza},
  \citenamefont {Lesiuk}, \citenamefont {Przybytek}, \citenamefont {Musia\l{}},
  \citenamefont {Moszynski}, \citenamefont {Lewenstein},\ and\ \citenamefont
  {Massignan}}]{GriningPRA2015}%
  \BibitemOpen
  \bibfield  {author} {\bibinfo {author} {\bibfnamefont {T.}~\bibnamefont
  {Grining}}, \bibinfo {author} {\bibfnamefont {M.}~\bibnamefont {Tomza}},
  \bibinfo {author} {\bibfnamefont {M.}~\bibnamefont {Lesiuk}}, \bibinfo
  {author} {\bibfnamefont {M.}~\bibnamefont {Przybytek}}, \bibinfo {author}
  {\bibfnamefont {M.}~\bibnamefont {Musia\l{}}}, \bibinfo {author}
  {\bibfnamefont {R.}~\bibnamefont {Moszynski}}, \bibinfo {author}
  {\bibfnamefont {M.}~\bibnamefont {Lewenstein}}, \ and\ \bibinfo {author}
  {\bibfnamefont {P.}~\bibnamefont {Massignan}},\ }\href {\doibase
  10.1103/PhysRevA.92.061601} {\bibfield  {journal} {\bibinfo  {journal} {Phys.
  Rev. A}\ }\textbf {\bibinfo {volume} {92}},\ \bibinfo {pages} {061601}
  (\bibinfo {year} {2015}{\natexlab{b}})}\BibitemShut {NoStop}%
\bibitem [{\citenamefont {Schmitz}\ \emph {et~al.}(2013)\citenamefont
  {Schmitz}, \citenamefont {Kr\"onke}, \citenamefont {Cao},\ and\ \citenamefont
  {Schmelcher}}]{Schmitz2013Breathing}%
  \BibitemOpen
  \bibfield  {author} {\bibinfo {author} {\bibfnamefont {R.}~\bibnamefont
  {Schmitz}}, \bibinfo {author} {\bibfnamefont {S.}~\bibnamefont {Kr\"onke}},
  \bibinfo {author} {\bibfnamefont {L.}~\bibnamefont {Cao}}, \ and\ \bibinfo
  {author} {\bibfnamefont {P.}~\bibnamefont {Schmelcher}},\ }\href@noop {}
  {\bibfield  {journal} {\bibinfo  {journal} {Physical Review A}\ }\textbf
  {\bibinfo {volume} {88}},\ \bibinfo {pages} {043601} (\bibinfo {year}
  {2013})}\BibitemShut {NoStop}%
\bibitem [{\citenamefont {Blume}(2010)}]{Blume2010TwoThreeToMany}%
  \BibitemOpen
  \bibfield  {author} {\bibinfo {author} {\bibfnamefont {D.}~\bibnamefont
  {Blume}},\ }\href@noop {} {\bibfield  {journal} {\bibinfo  {journal}
  {Physics}\ }\textbf {\bibinfo {volume} {3}},\ \bibinfo {pages} {74} (\bibinfo
  {year} {2010})}\BibitemShut {NoStop}%
\bibitem [{\citenamefont {{Zinner}}\ and\ \citenamefont
  {{Jensen}}(2013)}]{zinner-jensen2013}%
  \BibitemOpen
  \bibfield  {author} {\bibinfo {author} {\bibfnamefont {N.~T.}\ \bibnamefont
  {{Zinner}}}\ and\ \bibinfo {author} {\bibfnamefont {A.~S.}\ \bibnamefont
  {{Jensen}}},\ }\href {\doibase 10.1088/0954-3899/40/5/053101} {\bibfield
  {journal} {\bibinfo  {journal} {Journal of Physics G Nuclear Physics}\
  }\textbf {\bibinfo {volume} {40}},\ \bibinfo {eid} {053101} (\bibinfo {year}
  {2013})},\ \Eprint {http://arxiv.org/abs/1303.7351} {arXiv:1303.7351
  [nucl-th]} \BibitemShut {NoStop}%
\bibitem [{\citenamefont {{Andersen}}\ \emph {et~al.}(2016)\citenamefont
  {{Andersen}}, \citenamefont {{Dehkharghani}}, \citenamefont {{Volosniev}},
  \citenamefont {{Lindgren}},\ and\ \citenamefont {{Zinner}}}]{AndersenAe2016}%
  \BibitemOpen
  \bibfield  {author} {\bibinfo {author} {\bibfnamefont {M.~E.~S.}\
  \bibnamefont {{Andersen}}}, \bibinfo {author} {\bibfnamefont {A.~S.}\
  \bibnamefont {{Dehkharghani}}}, \bibinfo {author} {\bibfnamefont {A.~G.}\
  \bibnamefont {{Volosniev}}}, \bibinfo {author} {\bibfnamefont {E.~J.}\
  \bibnamefont {{Lindgren}}}, \ and\ \bibinfo {author} {\bibfnamefont {N.~T.}\
  \bibnamefont {{Zinner}}},\ }\href@noop {} {\bibfield  {journal} {\bibinfo
  {journal} {Scientific Reports}\ }\textbf {\bibinfo {volume} {6}},\ \bibinfo
  {pages} {28362} (\bibinfo {year} {2016})}\BibitemShut {NoStop}%
\bibitem [{\citenamefont {Girardeau}(1960)}]{girardeauJoMP1960}%
  \BibitemOpen
  \bibfield  {author} {\bibinfo {author} {\bibfnamefont {M.}~\bibnamefont
  {Girardeau}},\ }\href {\doibase http://dx.doi.org/10.1063/1.1703687}
  {\bibfield  {journal} {\bibinfo  {journal} {Journal of Mathematical Physics}\
  }\textbf {\bibinfo {volume} {1}},\ \bibinfo {pages} {516} (\bibinfo {year}
  {1960})}\BibitemShut {NoStop}%
\bibitem [{\citenamefont {Girardeau}\ and\ \citenamefont
  {Olshanii}(2004)}]{girardeauPRA2004}%
  \BibitemOpen
  \bibfield  {author} {\bibinfo {author} {\bibfnamefont {M.~D.}\ \bibnamefont
  {Girardeau}}\ and\ \bibinfo {author} {\bibfnamefont {M.}~\bibnamefont
  {Olshanii}},\ }\href {\doibase 10.1103/PhysRevA.70.023608} {\bibfield
  {journal} {\bibinfo  {journal} {Phys. Rev. A}\ }\textbf {\bibinfo {volume}
  {70}},\ \bibinfo {pages} {023608} (\bibinfo {year} {2004})}\BibitemShut
  {NoStop}%
\bibitem [{\citenamefont {Girardeau}\ and\ \citenamefont
  {Minguzzi}(2007)}]{girardeauPRL2007}%
  \BibitemOpen
  \bibfield  {author} {\bibinfo {author} {\bibfnamefont {M.~D.}\ \bibnamefont
  {Girardeau}}\ and\ \bibinfo {author} {\bibfnamefont {A.}~\bibnamefont
  {Minguzzi}},\ }\href {\doibase 10.1103/PhysRevLett.99.230402} {\bibfield
  {journal} {\bibinfo  {journal} {Phys. Rev. Lett.}\ }\textbf {\bibinfo
  {volume} {99}},\ \bibinfo {pages} {230402} (\bibinfo {year}
  {2007})}\BibitemShut {NoStop}%
\bibitem [{\citenamefont {Dehkharghani}\ \emph {et~al.}(2016)\citenamefont
  {Dehkharghani}, \citenamefont {Volosniev},\ and\ \citenamefont
  {Zinner}}]{DehkharghaniJoPBAMaOP2016}%
  \BibitemOpen
  \bibfield  {author} {\bibinfo {author} {\bibfnamefont {A.~S.}\ \bibnamefont
  {Dehkharghani}}, \bibinfo {author} {\bibfnamefont {A.~G.}\ \bibnamefont
  {Volosniev}}, \ and\ \bibinfo {author} {\bibfnamefont {N.~T.}\ \bibnamefont
  {Zinner}},\ }\href {http://stacks.iop.org/0953-4075/49/i=8/a=085301}
  {\bibfield  {journal} {\bibinfo  {journal} {Journal of Physics B: Atomic,
  Molecular and Optical Physics}\ }\textbf {\bibinfo {volume} {49}},\ \bibinfo
  {pages} {085301} (\bibinfo {year} {2016})}\BibitemShut {NoStop}%
\bibitem [{\citenamefont {Dehkharghani}\ \emph
  {et~al.}(2015{\natexlab{a}})\citenamefont {Dehkharghani}, \citenamefont
  {Volosniev}, \citenamefont {Lindgren}, \citenamefont {Rotureau},
  \citenamefont {Forss{\'e}n}, \citenamefont {Fedorov}, \citenamefont
  {Jensen},\ and\ \citenamefont {Zinner}}]{DehkharghaniSR2015}%
  \BibitemOpen
  \bibfield  {author} {\bibinfo {author} {\bibfnamefont {A.}~\bibnamefont
  {Dehkharghani}}, \bibinfo {author} {\bibfnamefont {A.}~\bibnamefont
  {Volosniev}}, \bibinfo {author} {\bibfnamefont {J.}~\bibnamefont {Lindgren}},
  \bibinfo {author} {\bibfnamefont {J.}~\bibnamefont {Rotureau}}, \bibinfo
  {author} {\bibfnamefont {C.}~\bibnamefont {Forss{\'e}n}}, \bibinfo {author}
  {\bibfnamefont {D.}~\bibnamefont {Fedorov}}, \bibinfo {author} {\bibfnamefont
  {A.}~\bibnamefont {Jensen}}, \ and\ \bibinfo {author} {\bibfnamefont
  {N.}~\bibnamefont {Zinner}},\ }\href {http://dx.doi.org/10.1038/srep10675}
  {\bibfield  {journal} {\bibinfo  {journal} {Scientific Reports}\ }\textbf
  {\bibinfo {volume} {5}},\ \bibinfo {pages} {10675 EP } (\bibinfo {year}
  {2015}{\natexlab{a}})},\ \bibinfo {note} {article}\BibitemShut {NoStop}%
\bibitem [{\citenamefont {Dehkharghani}\ \emph
  {et~al.}(2015{\natexlab{b}})\citenamefont {Dehkharghani}, \citenamefont
  {Volosniev},\ and\ \citenamefont {Zinner}}]{DehkharghaniPRA2015}%
  \BibitemOpen
  \bibfield  {author} {\bibinfo {author} {\bibfnamefont {A.~S.}\ \bibnamefont
  {Dehkharghani}}, \bibinfo {author} {\bibfnamefont {A.~G.}\ \bibnamefont
  {Volosniev}}, \ and\ \bibinfo {author} {\bibfnamefont {N.~T.}\ \bibnamefont
  {Zinner}},\ }\href {\doibase 10.1103/PhysRevA.92.031601} {\bibfield
  {journal} {\bibinfo  {journal} {Phys. Rev. A}\ }\textbf {\bibinfo {volume}
  {92}},\ \bibinfo {pages} {031601} (\bibinfo {year}
  {2015}{\natexlab{b}})}\BibitemShut {NoStop}%
\bibitem [{\citenamefont {Lehoucq}\ \emph {et~al.}(1998)\citenamefont
  {Lehoucq}, \citenamefont {Sorensen},\ and\ \citenamefont
  {Yang}}]{lehoucq1998arpack}%
  \BibitemOpen
  \bibfield  {author} {\bibinfo {author} {\bibfnamefont {R.~B.}\ \bibnamefont
  {Lehoucq}}, \bibinfo {author} {\bibfnamefont {D.~C.}\ \bibnamefont
  {Sorensen}}, \ and\ \bibinfo {author} {\bibfnamefont {C.}~\bibnamefont
  {Yang}},\ }\href@noop {} {\emph {\bibinfo {title} {ARPACK users' guide:
  solution of large-scale eigenvalue problems with implicitly restarted Arnoldi
  methods}}}\ (\bibinfo  {publisher} {SIAM},\ \bibinfo {year}
  {1998})\BibitemShut {NoStop}%
\bibitem [{\citenamefont {Sowi{\'n}ski}\ \emph {et~al.}(2013)\citenamefont
  {Sowi{\'n}ski}, \citenamefont {Grass}, \citenamefont {Dutta},\ and\
  \citenamefont {Lewenstein}}]{SowinskiGrass2013FewInteracting}%
  \BibitemOpen
  \bibfield  {author} {\bibinfo {author} {\bibfnamefont {T.}~\bibnamefont
  {Sowi{\'n}ski}}, \bibinfo {author} {\bibfnamefont {T.}~\bibnamefont {Grass}},
  \bibinfo {author} {\bibfnamefont {O.}~\bibnamefont {Dutta}}, \ and\ \bibinfo
  {author} {\bibfnamefont {M.}~\bibnamefont {Lewenstein}},\ }\href {\doibase
  10.1103/PhysRevA.88.033607} {\bibfield  {journal} {\bibinfo  {journal} {Phys.
  Rev. A}\ }\textbf {\bibinfo {volume} {88}},\ \bibinfo {pages} {033607}
  (\bibinfo {year} {2013})}\BibitemShut {NoStop}%
\bibitem [{\citenamefont {Sowi{\'n}ski}\ \emph {et~al.}(2015)\citenamefont
  {Sowi{\'n}ski}, \citenamefont {Gajda},\ and\ \citenamefont
  {Rzazewski}}]{SowinskiEEL2015}%
  \BibitemOpen
  \bibfield  {author} {\bibinfo {author} {\bibfnamefont {T.}~\bibnamefont
  {Sowi{\'n}ski}}, \bibinfo {author} {\bibfnamefont {M.}~\bibnamefont {Gajda}},
  \ and\ \bibinfo {author} {\bibfnamefont {K.}~\bibnamefont {Rzazewski}},\
  }\href {http://stacks.iop.org/0295-5075/109/i=2/a=26005} {\bibfield
  {journal} {\bibinfo  {journal} {EPL (Europhysics Letters)}\ }\textbf
  {\bibinfo {volume} {109}},\ \bibinfo {pages} {26005} (\bibinfo {year}
  {2015})}\BibitemShut {NoStop}%
\bibitem [{\citenamefont {Sowi{\'n}ski}(2015)}]{sowinski2015slightly}%
  \BibitemOpen
  \bibfield  {author} {\bibinfo {author} {\bibfnamefont {T.}~\bibnamefont
  {Sowi{\'n}ski}},\ }\href@noop {} {\bibfield  {journal} {\bibinfo  {journal}
  {Few-Body Systems}\ }\textbf {\bibinfo {volume} {56}},\ \bibinfo {pages}
  {659} (\bibinfo {year} {2015})}\BibitemShut {NoStop}%
\bibitem [{\citenamefont {P{\k e}cak}\ \emph {et~al.}(2016)\citenamefont {P{\k
  e}cak}, \citenamefont {Gajda},\ and\ \citenamefont
  {Sowi{\'n}ski}}]{Pecak2016Separation}%
  \BibitemOpen
  \bibfield  {author} {\bibinfo {author} {\bibfnamefont {D.}~\bibnamefont {P{\k
  e}cak}}, \bibinfo {author} {\bibfnamefont {M.}~\bibnamefont {Gajda}}, \ and\
  \bibinfo {author} {\bibfnamefont {T.}~\bibnamefont {Sowi{\'n}ski}},\ }\href
  {http://stacks.iop.org/1367-2630/18/i=1/a=013030} {\bibfield  {journal}
  {\bibinfo  {journal} {New Journal of Physics}\ }\textbf {\bibinfo {volume}
  {18}},\ \bibinfo {pages} {013030} (\bibinfo {year} {2016})}\BibitemShut
  {NoStop}%
\bibitem [{\citenamefont {P{\k e}cak}\ and\ \citenamefont
  {Sowi{\'n}ski}(2016)}]{Pecak2016Transition}%
  \BibitemOpen
  \bibfield  {author} {\bibinfo {author} {\bibfnamefont {D.}~\bibnamefont {P{\k
  e}cak}}\ and\ \bibinfo {author} {\bibfnamefont {T.}~\bibnamefont
  {Sowi{\'n}ski}},\ }\href@noop {} {\bibfield  {journal} {\bibinfo  {journal}
  {Phys. Rev. A}\ }\textbf {\bibinfo {volume} {94}},\ \bibinfo {pages} {042118}
  (\bibinfo {year} {2016})}\BibitemShut {NoStop}%
\bibitem [{\citenamefont {{Rotureau}}(2013)}]{RotureauEPJD2013}%
  \BibitemOpen
  \bibfield  {author} {\bibinfo {author} {\bibfnamefont {J.}~\bibnamefont
  {{Rotureau}}},\ }\href {\doibase 10.1140/epjd/e2013-40156-8} {\bibfield
  {journal} {\bibinfo  {journal} {European Physical Journal D}\ }\textbf
  {\bibinfo {volume} {67}},\ \bibinfo {eid} {153} (\bibinfo {year}
  {2013})}\BibitemShut {NoStop}%
\bibitem [{\citenamefont {Lindgren}\ \emph {et~al.}(2014)\citenamefont
  {Lindgren}, \citenamefont {Rotureau}, \citenamefont {Forss\'en},
  \citenamefont {Volosniev},\ and\ \citenamefont {Zinner}}]{LindgrenNJoP2014}%
  \BibitemOpen
  \bibfield  {author} {\bibinfo {author} {\bibfnamefont {E.~J.}\ \bibnamefont
  {Lindgren}}, \bibinfo {author} {\bibfnamefont {J.}~\bibnamefont {Rotureau}},
  \bibinfo {author} {\bibfnamefont {C.}~\bibnamefont {Forss\'en}}, \bibinfo
  {author} {\bibfnamefont {A.~G.}\ \bibnamefont {Volosniev}}, \ and\ \bibinfo
  {author} {\bibfnamefont {N.~T.}\ \bibnamefont {Zinner}},\ }\href
  {http://stacks.iop.org/1367-2630/16/i=6/a=063003} {\bibfield  {journal}
  {\bibinfo  {journal} {New Journal of Physics}\ }\textbf {\bibinfo {volume}
  {16}},\ \bibinfo {pages} {063003} (\bibinfo {year} {2014})}\BibitemShut
  {NoStop}%
\bibitem [{\citenamefont {Hu}\ \emph {et~al.}(2016)\citenamefont {Hu},
  \citenamefont {Guan},\ and\ \citenamefont {Chen}}]{HuNJoP2016}%
  \BibitemOpen
  \bibfield  {author} {\bibinfo {author} {\bibfnamefont {H.}~\bibnamefont
  {Hu}}, \bibinfo {author} {\bibfnamefont {L.}~\bibnamefont {Guan}}, \ and\
  \bibinfo {author} {\bibfnamefont {S.}~\bibnamefont {Chen}},\ }\href
  {http://stacks.iop.org/1367-2630/18/i=2/a=025009} {\bibfield  {journal}
  {\bibinfo  {journal} {New Journal of Physics}\ }\textbf {\bibinfo {volume}
  {18}},\ \bibinfo {pages} {025009} (\bibinfo {year} {2016})}\BibitemShut
  {NoStop}%
\bibitem [{\citenamefont {Deuretzbacher}\ \emph {et~al.}(2016)\citenamefont
  {Deuretzbacher}, \citenamefont {Becker}, \citenamefont {Bjerlin},
  \citenamefont {Reimann},\ and\ \citenamefont
  {Santos}}]{deuretzbacher2016spin}%
  \BibitemOpen
  \bibfield  {author} {\bibinfo {author} {\bibfnamefont {F.}~\bibnamefont
  {Deuretzbacher}}, \bibinfo {author} {\bibfnamefont {D.}~\bibnamefont
  {Becker}}, \bibinfo {author} {\bibfnamefont {J.}~\bibnamefont {Bjerlin}},
  \bibinfo {author} {\bibfnamefont {S.}~\bibnamefont {Reimann}}, \ and\
  \bibinfo {author} {\bibfnamefont {L.}~\bibnamefont {Santos}},\ }\href@noop {}
  {\bibfield  {journal} {\bibinfo  {journal} {arXiv preprint arXiv:1611.04418}\
  } (\bibinfo {year} {2016})}\BibitemShut {NoStop}%
\bibitem [{\citenamefont {{Dehkharghani}}\ \emph {et~al.}(2017)\citenamefont
  {{Dehkharghani}}, \citenamefont {{Bellotti}},\ and\ \citenamefont
  {{Zinner}}}]{Dehkharghani2017}%
  \BibitemOpen
  \bibfield  {author} {\bibinfo {author} {\bibfnamefont {A.~S.}\ \bibnamefont
  {{Dehkharghani}}}, \bibinfo {author} {\bibfnamefont {F.~F.}\ \bibnamefont
  {{Bellotti}}}, \ and\ \bibinfo {author} {\bibfnamefont {N.~T.}\ \bibnamefont
  {{Zinner}}},\ }\href@noop {} {\bibfield  {journal} {\bibinfo  {journal}
  {ArXiv e-prints}\ } (\bibinfo {year} {2017})},\ \Eprint
  {http://arxiv.org/abs/1703.01836} {arXiv:1703.01836 [cond-mat.quant-gas]}
  \BibitemShut {NoStop}%
\end{thebibliography}%
\end{document}